\documentclass[a4paper,11pt]{article}
\pdfoutput=1 % if your are submitting a pdflatex (i.e. if you have
\usepackage{jheppub} % for details on the use of the package, please
\usepackage[T1]{fontenc} % if needed
\usepackage{caption}

\numberwithin{equation}{section}
\numberwithin{table}{section}

\newcommand{\beq}{\begin{equation}}  \newcommand{\eeq}{\end{equation}}
\newcommand{\bal}{\begin{aligned}}   \newcommand{\eal}{\end{aligned}}
\newcommand{\bea}{\begin{eqnarray}}  \newcommand{\eea}{\end{eqnarray}}

\newcommand{\bmat}{\left(\begin{array}}
\newcommand{\emat}{\end{array}\right)}

\newcommand{\nn}{\nonumber}

\newcommand{\cO}{\mathcal{O}}

\newcommand{\cK}{\mathcal{K}}

\newcommand{\cW}{\mathcal{W}}
\newcommand{\cG}{\mathcal{G}}

\newcommand{\I}{\text{i}}

\usepackage{cancel}
\usepackage[normalem]{ulem}

\newcommand{\be}{\begin{equation}}
\newcommand{\ee}{\end{equation}}

\usepackage{amsthm}

\newcommand{\kom}{\, ,\quad }

\newcommand*{\dif}{{\,\rm d}}
\newcommand*{\p}{\mathop{}\!\mathrm \partial}

\newcommand*{\lm}{L_{\text{max}}}
\newcommand*{\gs}{g_{\text{s}}}

\newcommand{\re}{\text{Re}}
\newcommand{\im}{\text{Im}}

\title{Searching the Landscape of Flux Vacua with Genetic Algorithms}
\author{Alex Cole,}
\author{Andreas Schachner,}
\author{Gary Shiu}

\affiliation{Department of Physics, University of Wisconsin, 1150 University Avenue, Madison, WI 53706, USA}
\emailAdd{acole4@wisc.edu}
\emailAdd{schachner@thphys.uni-heidelberg.de}
\emailAdd{shiu@physics.wisc.edu}

\abstract{In this paper, we employ genetic algorithms to explore the landscape of type IIB flux vacua. We show that genetic algorithms can efficiently scan the landscape for viable solutions satisfying various criteria. More specifically, we consider a symmetric $T^{6}$ as well as the conifold region of a Calabi-Yau hypersurface. We argue that in both cases genetic algorithms are powerful tools for finding flux vacua with interesting phenomenological properties. We also compare genetic algorithms to algorithms based on different breeding mechanisms as well as random walk approaches.}


\arxivnumber{1907.10072}

\begin{document}
\maketitle
\tableofcontents

%%%%%%%%%%%%%%%%%%%%%%%%%%%%%%%%%%%%%%%%%%%%%%%
\section{Introduction}
%%%%%%%%%%%%%%%%%%%%%%%%%%%%%%%%%%%%%%%%%%%%%%%

Since the emergence of the string theory landscape \cite{Bousso:2000xa,Susskind:2003kw,Douglas:2003um}, its predictions (and even the existence of phenomenologically viable vacua) have remained a matter of debate \cite{Banks:2003es,Banks:2004xh,Banks:2012hx}. Over the years, several approaches including systematic scans of explicit constructions and statistical analyses have been considered for exploring the landscape \cite{Ashok:2003gk,Denef:2004ze,Douglas:2004zu,Denef:2004cf,Susskind:2004uv,Douglas:2004qg,Dine:2004is,Conlon:2004ds,Kallosh:2004yh,Marchesano:2004yn,Dine:2005yq,Acharya:2005ez,Dienes:2006ut,Gmeiner:2005vz,Douglas:2006xy}. The heart of the problem is that the enormous number of vacua in the landscape makes a straightforward scan computationally infeasible. The discrete landscape can be viewed as the complement of a continuum of seemingly consistent low-energy effective field theories (EFTs) that cannot descend from a string compactification, deemed the swampland \cite{Vafa:2005ui,Ooguri:2006in}, see \cite{Palti:2019pca} for a review. While the latter has received much attention in recent years, progress in data science might allow for systematic studies of the landscape itself, as demonstrated with a variety of techniques such as topological data analysis \cite{Cole:2018emh} and machine learning \cite{He:2017aed, Krefl:2017yox, Ruehle:2017mzq, Carifio:2017bov, Wang:2018rkk, Bull:2018uow, Klaewer:2018sfl, Mutter:2018sra,Halverson:2019tkf,He:2019vsj}.

In this paper, we study the landscape of string vacua using \emph{Genetic Algorithms} (GAs) \cite{Holland1975,David1989,Holland1992,Reeves2002,haupt,Michalewicz2004}. GAs search for optimal solutions to problems with huge input spaces via a natural selection process. The population of potential solutions evolves according to the fitness of individual members. 
Each individual is specified by its genotype, a string of data that defines the member's location in the input space.
An individual's phenotype characterizes its properties as a solution, generally summarized by some fitness function.  Via dynamics motivated by natural selection, genetic information related to the fittest members propagates to subsequent generations. Overall, the population adapts to the extrinsic factors implemented by the fitness.

In the past few decades, genetic algorithms have proven useful in diverse areas of physics including particle phenomenology \cite{Allanach:2004my,Akrami:2009hp,Abel:2018ekz}, astrophysics \cite{Metcalfe:2000xn,Mokiem:2005qf,Rajpaul:2012wu} and cosmology \cite{Nesseris:2012tt,Hogan:2014qsa}. Applications of GAs to the landscape have so far been rather limited \cite{Blaback:2013ht,Damian:2013dq,Damian:2013dwa,Blaback:2013fca,Blaback:2013qza,Abel:2014xta,Ruehle:2017mzq}. This is surprising given that they are very successful in scanning large data sets for viable solutions. 
Moreover, for deterministic algorithms, difficulties due to the large size of the landscape are exacerbated by computational complexity, as can be seen 
in several toy models \cite{Denef:2006ad,Bao:2017thx,Denef:2017cxt,Halverson:2018cio}. 
We might hope to avoid some of these issues using stochastic search methods based on natural selection processes, such as GAs. {On the deep learning side, a closely related technique is Reinforcement Learning (RL). GAs and RL are both used to train machine agents to solve complex tasks in uncertain environments \cite{silver2016mastering,2017arXiv170303864S}.}

Ultimately, we are interested in selecting vacua 
based on 
their phenomenological properties such as
  the cosmological constant or
certain desirable low energy spectra and couplings.
As a first step, we focus in this paper 
on three such quantities: the string coupling, the superpotential value in the stabilized vacua, and the moduli masses.
 Nonetheless, it is straightforward to extend our GA analysis to other contexts.
 Methods along the lines outlined in this paper can be used e.g. to
 test the WGC \cite{ArkaniHamed:2006dz} (and its various strong forms \cite{Brown:2015iha,Brown:2015lia,Heidenreich:2015nta,Heidenreich:2016aqi,Montero:2016tif,Andriolo:2018lvp}).
 As a proof of concept,
 the problems addressed in this paper are not too computationally intensive, so all the algorithms we used
  can be implemented in Mathematica.
 However, it is worth mentioning that there are 
 many publicly available packages for GAs written for different platforms,
  see e.g. \cite{1995ApJS..101..309C,charbonneau1995user,charbonneau2002introduction,charbonneau2002release}\footnote{For a maintained list of packages for genetic programming in general, see \href{http://geneticprogramming.com/software/}{http://geneticprogramming.com/software/}.}.
  
  \begin{figure}[t!]
  \centering
  \begin{tabular}{c|c}
  Genetic Algorithms & Landscape \\ 
  \hline 
  \hline 
  individuals & flux vacua\\ 
  \hline 
  chromosome & flux vector \\ 
  \hline 
  alleles & fluxes \\ 
  \hline 
   phenotype & masses, couplings, etc. \\ 
  \hline 
  fitness & function of masses, couplings, etc. \\ 
  \hline 
  boundary conditions & SUSY condition, gauge fixing, tadpole \\ 
  \end{tabular} 
  \captionof{table}{Dictionary relating terms used in the context of GAs and the flux landscape}\label{tab:1} 
  \end{figure}

In this paper, we explore the type IIB flux landscape, for which our genetic algorithms are (mostly) based on the following identifications, see also Tab.~\ref{tab:1}. The members of a population are distinct string vacua, each determined by a choice of (integer-quantized) fluxes. In other words, the chromosomes are flux vectors of maximal length $4(h^{(2,1)}+1)$ characterizing the genotype of a vacuum. Solving the F-term constraints for these fluxes determines the VEVs of the various moduli and gives the phenotype of a vacuum. The corresponding fitness is a function of physical observables including masses and couplings. Parent vacua are selected according to their fitness, and breed pairwise to construct child vacua. Parents breed via a crossover procedure where parts of one chromosome are replaced by the corresponding parts of a mate's chromosome. Additionally, mutation modifies some fluxes at random. After solving the F-flatness conditions and checking gauge inequivalence and the size of tadpoles, we can define a new population by choosing an appropriate number of child vacua as our subsequent generation.

Repeating the above steps for many generations results in a population gathering around a point of maximal fitness, i.e., vacua satisfying physical conditions specified by the fitness. Moreover, we find that GAs are significantly more efficient at finding physical solutions than randomly choosing fluxes. In other words, GAs are inherently superior to brute force approaches. This is a hint that the algorithm exploits some underlying structure of the flux landscape that is unknown at the outset of the search. This learning of structure can be confirmed by studying the ``shapes'' of individual generations using persistent homology, a technique previously applied to the landscape in \cite{Cole:2018emh}. {Exploitation of this structure leads to improved efficiency of a GA compared to simulated annealing, see Sect. \ref{sec:Metro}.}

GAs, or stochastic search algorithms in general, are not effective for ``needle-in-a-haystack'' type problems. These problems do not allow for a distinction between incorrect solutions. In other words, all incorrect solutions are equally bad, so the performance of a GA is reduced to that of a random scan. Turning this around, it is important that a GA is searching a \emph{fitness landscape} that is pseudo-continuous. Stated differently, there should exist a neighborhood around the optimal solution in which the fitness is well-behaved. This can be quantified by considering the \emph{fitness-distance correlation} \cite{jones1995fitness,collard1998fitness} with distance measured in the input space. 

{In general, stochastic search algorithms are efficient when the fitness landscape exhibits a funnel-like topography (see \cite{bryngelson1995funnels}, and recently in the context of the landscape \cite{Khoury:2019yoo}), so that the optimal solution can be approached via small steps. This encodes itself in a large and negative fitness-distance correlation near the optimal solution, see Sect. \ref{sec:ConvEv}. Strictly speaking, the viability of a given search algorithm is determined empirically. If a solution or near-solution is found, the algorithm may be considered successful. If a solution is not found, either it does not exist, or it is not readily accessible to the algorithm at hand. In the latter case, computing the fitness-distance correlation can demonstrate the lack of a favorable funnel topography. This can sometimes be ameliorated by an alternative choice of problem encoding \cite{ruml1996easily}, but in general finding an efficient encoding requires solving the problem itself (in which case no search algorithm is necessary) or finding successful deterministic heuristics, which we aim to avoid due to its problem-specific nature.}

In the presence of a fitness landscape the flow of the population shows a strong pull on certain observables so that the population quickly gathers around suitable solutions. This behavior typically depends on the precise definition of the fitness itself, especially when considering several search parameters. In contrast, ``needle-in-a-haystack'' situations can only randomly sample the space of solutions since unwanted solutions cannot be distinguished. As we will see, all models considered in this paper do not belong to the class of a ``needle-in-a-haystack''. It has previously been argued that due to the structure and correlations within the landscape GAs are expected to work well more generally \cite{Abel:2014xta}. It seems therefore promising to us that GAs constitute valuable tools to systematically study the landscape.

This paper is organized as follows. First, we review the construction of flux vacua in type IIB compactifications in Section \ref{sec:Flux}. We then collect the basic principles of genetic algorithms in Section \ref{sec:GA}. Subsequently, we define an algorithm suitable for studying type IIB flux compactifications. Afterwards, we apply our algorithm to two examples: a hypersurface in the weighted projective space ${\bf WP}^{4}_{1,1,1,1,4}$ (Section \ref{sec:Hyper}) and the symmetric $T^{6}$ (Section \ref{sec:T6}). We conclude in Section \ref{sec:summary}.

%%%%%%%%%%%%%%%%%%%%%%%%%%%%%%%%%%%%%%%%%%%%%%%
\section{Flux compactifications}\label{sec:Flux}
%%%%%%%%%%%%%%%%%%%%%%%%%%%%%%%%%%%%%%%%%%%%%%%

In this section, we briefly review type IIB flux compactifications on Calabi-Yau orientifolds in the presence of background fluxes, see \cite{Grana:2005jc,Douglas:2006es} for reviews. In these setups, 3-form fluxes stabilize the axio-dilaton and complex structure moduli. Due to flux quantization and tadpole cancellation, the resulting space of vacuum solutions is a discretuum distributed over the moduli space \cite{Bousso:2000xa,Giddings:2001yu}.

We follow the conventions of \cite{DeWolfe:2004ns}. Consider a Calabi-Yau threefold $M$ with $h^{(2,1)}$ complex structure moduli. We take a symplectic basis $\{A^a,B_b\}$ for the $b_3=2h^{(2,1)}+2$ three-cycles, with $a,b=1,\dots,h^{(2,1)}+1$. The dual cohomology elements $\alpha_a,\beta^b$ satisfy
\begin{align}
	\int_{A^a}\alpha_b=\delta^a_b,\quad \int_{B_b}\beta^a=-\delta^a_b,\quad\int_M \alpha_a\wedge \beta^b=\delta^b_a\, .
\end{align}
From the unique holomorphic three-form $\Omega$, we define the periods $z^a\equiv \int_{A^{a}}\Omega,~\mathcal{G}_b\equiv\int_{B_b}\Omega$, which form the $b_3$-vector $\Pi(z)\equiv(\mathcal{G}_b,z^a)$. Additionally
\begin{align}
	\int_M\Omega\wedge\overline{\Omega}=\overline{z}^a\mathcal{G}_a-z^a\overline{\mathcal{G}}_a=-\Pi^\dag\cdot\Sigma\cdot\Pi
\end{align}
in terms of the symplectic matrix
\begin{align}
 	\Sigma=\left(\begin{matrix}0&1\\ -1&0\end{matrix}\right)
\end{align} 
whose entries are $(h^{(2,1)}+1)\times(h^{(2,1)}+1)$ matrices. The RR and NSNS 3-form fluxes are quantized and can be expanded in the $\alpha,\beta$ basis as
\begin{align}
  F_3=-(2\pi)^2\alpha^{\prime}(f_a\alpha_a+f_{a+h^{(2,1)}+1}\beta^a),\quad H_3=-(2\pi)^2\alpha^{\prime}(h_a\alpha_a+h_{a+h^{(2,1)}+1}\beta^a)\, .
\end{align}
Here, $f$ and $h$ are two integer-valued $b_3$-vectors. From now on we set $(2\pi)^2\alpha^{\prime}=1$ and define
\begin{equation}\label{eq:Flux} 
\mathbf{N}=\left (f_1,\ldots,f_{2h^{(2,1)}+2},h_{1},\ldots,h_{2h^{(2,1)}+2}\right )^{T}
\end{equation}
for later purposes. The fluxes induce a superpotential for the complex structure moduli and axio-dilaton $\phi\equiv C_0+\I e^{-\varphi}$, given by \cite{Gukov:1999ya}
\begin{align}
  \cW=\int_M G_3\wedge \Omega(z)=(f-\phi h)\cdot \Pi(z)\, .
\end{align}
In type IIB supergravity the $3$-form fluxes only appear in the combination $G_3\equiv F_3-\phi H_3$.

The tree-level $\mathcal{N}=1$ F-term scalar potential induced by $3$-form fluxes is of no-scale type, given by
\begin{equation}
V=\mathrm{e}^{\cK}\left (\cK^{a\bar{b}}D_{a}\cW\, D_{\bar{b}}\overline{\cW}+\cK^{\phi\bar{\phi}}D_{\phi}\cW\, D_{\bar{\phi}}\overline{\cW}\right )\, .
\end{equation}
Here, $\cK^{a\bar{b}}$ is the inverse K{\"a}hler metric on complex structure moduli space and $D_{a}\cW=(\p_{z_{a}}+(\p_{z_{a}}\cK))\cW$ the associated K{\"a}hler derivative (similarly for $\phi$). The corresponding mass matrix for the real scalar fields is given by
\begin{equation}\label{eq:MassMatrix} 
M_{IJ}=\p_{I}\p_{J}V
\end{equation}
evaluated at a minimum of $V$ with $I,J\in\lbrace\re(z_{a}),\im(z_{a}),\re(\phi),\im(\phi)\rbrace$.

We are interested in vacua with vanishing F-terms
\begin{align}\label{eqn:fflat1}
  D_\phi \cW&=\frac{1}{\overline{\phi}-\phi}(f-\overline{\phi}h)\cdot \Pi(z)=0\, ,\\\label{eqn:fflat2}
  D_a \cW&= (f-\phi h)\cdot(\partial_a \Pi(z)+\Pi(z)\partial_a\mathcal{K})=0
\end{align} 
The K\"ahler potential for the axio-dilaton and complex structure moduli is given by
\begin{align}
  \mathcal{K}&=-\log\left(\I\int_M\Omega\wedge\overline{\Omega}\right)-\log\left(-\I(\phi-\overline{\phi})\right)\nonumber\\
  &=-\log(-\I\Pi^\dag\cdot\Sigma\cdot \Pi)-\log(-\I(\phi-\overline{\phi}))
\end{align}
The F-flatness conditions (\ref{eqn:fflat1}) and (\ref{eqn:fflat2}) imply that the (3,0) and (1,2) parts of the fluxes vanish, so that $G_3$ is imaginary self-dual, $\star_6 G_3=\I G_3$. 

The D3-brane charge induced by the fluxes can be written as
\begin{align}
  N_{\rm flux}=\int_M F_3\wedge H_3=f\cdot\Sigma\cdot h\, .
\end{align}
One can show that $N_{\rm flux}>0$ for imaginary self-dual fluxes. Hence, negative D3-brane charges have to appear to ensure tadpole cancellation. These negative charges can be induced by orientifolding. In this paper, we will not be explicit about orientifolding. We will rather take $L_{\rm max}$ as an adjustable parameter and, along the lines of \cite{Ashok:2003gk,Denef:2004ze,DeWolfe:2004ns}, consider solutions with
\begin{align}\label{eqn:tadpole}
  0<N_{\rm flux}\leq L_{\rm max}\, 
\end{align}
with remaining charge cancelled by mobile D3-branes. As shown in \cite{Cole:2018emh}, $L_{\rm max}$ sets the scale at which interesting structure in the moduli space distribution appears.

It is crucial that we fix gauge symmetries relating equivalent vacua when running our GA. Otherwise, we might end up with redundancies in our population of vacua. In our case, the symmetry group is given by $\mathcal{G}=SL(2,\mathbb{Z})_\phi\times \Gamma$ where $SL(2,\mathbb{Z})_\phi$ is the S-duality group from type IIB and $\Gamma$ is the modular group acting on the complex structure moduli space. We choose a gauge-fixing prescription in which each vacuum is mapped to the corresponding fundamental domain where we have to keep track of all fluxes. Throughout, we will consider the evolution of stabilized VEVs for the axio-dilaton and complex structure moduli, and functions thereof.

%%%%%%%%%%%%%%%%%%%%%%%%%%%%%%%%%%%%%%%%%%%%%%%
\section{Genetic algorithms}\label{sec:GA}
%%%%%%%%%%%%%%%%%%%%%%%%%%%%%%%%%%%%%%%%%%%%%%%

In this section, we collect the necessary ingredients to study genetic algorithms in the context of string compactifications, see also \cite{Reeves2002,Allanach:2004my,Abel:2014xta} for more pedagogical introductions. We begin with a brief summary of GAs. Afterwards, we turn to applications to flux vacua and describe the algorithm applied in the remainder of this paper. We also highlight novel complications encountered when using GAs to search the landscape, namely that consistency conditions are not necessarily preserved by the breeding process. 
Finally, we elaborate further on the suitability and performance of GAs in the context of flux compactifications.

\subsection{Generalities}\label{sec:GAGen}

A GA in the sense of \cite{Holland1975,Holland1992} can be described as follows. One begins with a population of  $p$ randomly chosen individuals, each specified by a string of data referred to as the \emph{chromosome}. The chromosome describes the defining input parameters of each individual, thereby encoding the individual's \emph{genotype}. The chromosome consists of \emph{alleles} whose number and values depend on the model in question. Typically, for alleles taking values in a smaller range, a smaller population size can be chosen, see the discussion in Sect.~\ref{sec:GenAlg}. The physical characteristics of a member, also called its \emph{phenotype}, are obtained by computing functions of the genotype. In a physics context, this typically corresponds to a set of physical observables like masses, couplings, or representations. Now, the algorithm proceeds by repeating the following three steps.

First, the process of selection is introduced by declaring certain individuals to be more competitive for breeding than others. This is typically achieved by collecting \emph{pairs} where the single members in a pair are referred to as \emph{parents}. As in nature, specific individuals have physical properties distinguishing them from the typical individual and making them more likely to procreate. This preeminence can be taken into account by assigning to each member a \emph{fitness} based on the phenotype.

Several ways to define the fitness as well as various selection methods have been discussed in the literature, see e.g. \cite{baker1987reducing,goldberg1989genetic,whitley1989genitor,saliby1990descriptive,
goldberg1991comparative,miihlenbein1994science,hancock1994empirical,hancock1995selection,
blickle1996comparison,thierens1998selection,lohr1999sampling,rudolph2000takeover,
rudolph2001takeover,reeves2002genetic,haupt2004practical}. In this paper, we mostly focus on the so-called \emph{roulette-wheel selection}, but see Sect.\ \ref{sec:masses}. Here, one normalizes the fitness function to be a probability distribution. Pairs are constructed by choosing individuals according to this probability distribution. Note that this method of parent selection explicitly takes self-reproduction into account.

Secondly, individuals breed to construct a new population, made up of members called \emph{children}. This is achieved by splicing the two chromosomes of a single pair together. We typically employ a \emph{two-point crossover} where the parent chromosomes are both cut at the same two randomly selected positions, swapping the middle sections. Afterwards, we pick one of the two new chromosomes at random as the genotype of a new individual in the descendant population.

Thirdly, and this is the major feature responsible for the efficiency of genetic algorithms, the children's chromosomes are altered by \emph{mutation}. More specifically, for randomly selected children, a fraction of random alleles, typically about $1\%$, are modified. This is necessary to prevent the algorithm from stagnation. For example, the population can cluster around a local maximum of the fitness, even though a better global maximum is available. The algorithm might never gain knowledge of the latter without mutation. In this sense, mutation should not be viewed as a tool to increase convergence, but more crucially as an integral part of the algorithm itself. Along these lines, we will compare different mutation rates within a simple example in Sect.~\ref{sec:MutHyper}. Taking the population size $p$ to remain constant\footnote{Increasing $p$ dynamically during the evolution comes at the expense of increased computational effort. Moreover, the rate of convergence is not improved as long as we start with a large population, see also Sect.~\ref{sec:GenAlg}.}, the new population is obtained by taking all children, replacing the least fit individual found after mutation with the fittest individual from the previous population. This last step, called \emph{elitist selection}, forces the maximum fitness to increase monotonically with each generation.

The above three steps form a \emph{generation}. We repeat them several times, until a limiting rate of convergence is reached. In general, only a certain fraction of the population can reach the optimal solution, see e.g. the discussion in Sect.~1.2 of \cite{Abel:2014xta}.

\subsection{Genetic algorithms for Flux Vacua}\label{sec:GenAlg} 

In this section we translate the general notion of GAs into the language of type IIB flux compactifications. Throughout this section we employ a notation where $\langle \cO\rangle_{A}$ denotes the vacuum expectation value of some quantity $\cO$ computed by solving the F-flatness conditions for the fluxes $\mathbf{N}_{A}$.

A \emph{population} of size $p$ is a set of flux vacua $V_{A}$, $A=1,\ldots ,p$, obtained by solving the F-term constraints $D_{a}\cW=0=D_{\phi}\cW$, $a=1,\ldots,h^{(2,1)}$, and fixing the gauge redundancy. Apart from intrinsic geometric quantities specified by the choice of a compactification space, input parameters are only flux numbers $\mathbf{N}_{A}=\bigl (N_{A}^{1},\ldots, N_{A}^{2b_{3}}\bigl )^{T}$ (recall eq.~\eqref{eq:Flux}), i.e.,
\begin{equation}
V_{A}=V(\mathbf{N}_{A})=V\bigl (N_{A}^{1},\ldots, N_{A}^{2b_{3}}\bigl )\, .
\end{equation}
More specifically, $V_{A}$ encodes all elementary information about the vacuum which corresponds in our case to the VEVs of all moduli fields $\langle z_{a}\rangle_{A}$ and the axio-dilaton $\langle\phi\rangle_{A}$. Hence, each vacuum $V_{A}$ has physical attributes including
\begin{itemize}
\item VEVs $\langle z_{a}\rangle_{A}$, $\langle\phi\rangle_{A}$
\item the value of $N_{\text{flux}}(\mathbf{N}_{A})$
\item the value of $\cW_{0}(\mathbf{N}_{A})=\langle\cW\rangle_{A}$, the string coupling and the moduli masses.
\end{itemize}
These characteristics specify the vacuum's location in moduli space, the tadpole as well as various mass scales etc. Initially, we choose a certain number of fluxes at random, whereas fluxes of any descendant population are determined through the pairing process (see below). It is crucial that we restrict the tadpole such that $N_{\text{flux}}\leq L_{\text{max}}$, albeit we do not fix it to a certain value. Otherwise, the population might be driven towards solutions with arbitrarily large tadpole. {Restricting the size of our tadpole may present limits to the sorts of vacua accessible to our algorithm. For example, it has been noted that in some examples F-term constraints force vacua at weak coupling, large complex structure, and large volume to have large $N_{\rm flux}$ \cite{Betzler:2019kon}.}

The \emph{fitness} can be used to find vacua with certain values for the VEVs or functions thereof, such as $\cW_{0}$, which we collectively denote $\cO^{(i)}$ in this section. For a fixed compactification space, the fitness of some vacuum $V_{A}$ will be a function of the fluxes, $F_{A}=F\left (\mathbf{N}_{A}\right )$. Say we are interested in vacua with values $\cO^{(i),\ast}$ for some observables $\cO^{(i)}$. (Throughout this paper, we denote with the superscript $\,^{\ast}$ the optimal solution within our GA.) We write the value of the observable $\cO^{(i)}$ in the vacuum $V_{A}$ as $\cO^{(i)}_{A}=\langle\cO^{(i)}\rangle_{A}$. We define the associated fitness $F_{A}$ of $V_{A}$ as
\begin{equation}\label{eq:GenFitness} 
	F_{A}=\dfrac{1}{\mathcal{N}}\left [\sum_{i}\, w_{i} f_{i}(\cO^{(i)}_{A},\cO^{(i),\ast},\delta \cO^{(i)})+b\right ]\, .
\end{equation}
Here, we introduce weights $w_{i}$ and a normalisation factor $\mathcal{N}$ so that $\sum_{A}\, F_{A}=1$. We typically choose $f_{i}$ to be a Gaussian of width $\delta \cO^{(i)}$, but other choices also work well. The offset $b$ is necessary to prevent $F_A$ from localizing around a local maximum of one of the $f_i$. In general, we have to keep in mind that several of the $\cO^{(i)}$ could be correlated. Thus, it also has to be checked that the required values $\cO^{(i),\ast}$ are compatible with each other. Finally, the dynamics of the algorithm is generically highly sensitive to the choice of weights $w_i$.

Using the fitness, we generate a set of pairs $(V_{A},V_{B})$ of vacua. We allow the two vacua in a pair to be identical. We typically take some number $P> p$ of random pairs, large enough to guarantee enough gauge-inequivalent solutions after breeding satisfying tadpole constraints. This is a minor setback compared to the general algorithm described in Sect.~\ref{sec:GAGen}. There is typically no way to tell whether a pair leads to a viable solution a priori.  Still, random search algorithms face similar problems in this context, while they do not benefit from the crucial interplay of selection, crossover and mutation. Thus, we expect our algorithm to be more efficient than a Monte-Carlo procedure, notwithstanding the aforementioned difficulties.

A child vacuum $v_{C}$ is constructed from a pair $(V_{A},V_{B})$ in the following way. First, we construct a new string of fluxes along the lines of Sect.~\ref{sec:GAGen}. That is, the new genotype $\mathbf{n}_{C}=(n_{C}^{1},\ldots,n_{C}^{2b_{3}})^{T}$ is given by
\begin{equation}\label{eq:Child} 
\mathbf{n}_{C}=\left (N_{A}^{1},\ldots,N_{A}^{k_{1}},N_{B}^{k_{1}+1},\ldots, N_{B}^{k_{2}},N_{A}^{k_{2}+1},\ldots, N_{A}^{2b_{3}}\right )^{T}
\end{equation}
where $k_{1},k_{2}\in\lbrace 1,\ldots, 2b_{3}\rbrace$, $k_{1}\leq k_{2}$, are two randomly chosen integers. Mutations are applied by replacing a certain number of fluxes by some new random value. We introduce the mutation rate $q_{\text{mut}}$ and perform a mutation on $n_{\text{mut}}$ flux numbers of a single individual whenever $q<q_{\text{mut}}$ for a randomly chosen number $q\in[0,1]$.

Next, all newly obtained strings of fluxes are plugged into the F-term constraints to judge whether they meet all relevant criteria. More specifically, we have to check the existence of solutions, the size of the flux-induced D3-charge $N_{\text{flux}}$ as well as gauge inequivalence. Finally, the new genotype potentially results in a well-defined vacuum $v_{C}=V\left (\mathbf{n}_{C}\right )$.

A new population is defined as a random choice of $p$ children $v_{C}$. As mentioned above, it is hence mandatory that the previous step leads to more than $p$ inequivalent vacua by initially generating $P> p$ pairs. Having defined the population of the next generation, we can continue by computing the associated fitness of each member and going through the same steps again.

We also have to think about the initial population size. We typically restrict to tadpoles smaller than some flux scale $L_{\text{max}}$ and compute the relevant quantities. Since genetic algorithms are designed to find the global maximum of the fitness function, it is necessary to have access to all the parameter space. This means that every allele should include every possible flux choice. We restrict to fluxes in the interval $[-L_{\text{max}},L_{\text{max}}]$ for simplicity.\footnote{This is typically sufficient for the examples considered here since there are almost no solutions beyond these limits, cf. the discussion at the end of Sect.~4.3.2 and the second paragraph of Sect.~5.4 in \cite{DeWolfe:2004ns}.} If we take the population size to satisfy
\begin{equation}\label{eq:sizep}
	p\geq (2\lm+1)\, 4(h^{(2,1)}+1)\, ,
\end{equation}
then this is most certainly true. This is obviously a very naive estimate, but we expect it to be sufficient for the level of our discussion.

\subsection{Schemata and fitness distance correlation}\label{sec:ConvEv} 

We call the space parametrized by the $\cO_{i}$ the search space, in contrast to the parameter space of fluxes $\mathbf{N}$. A GA converges in search space, but not necessarily in parameter space. By this we mean that the final population is dominated by members for which $\cO^{(i)}_{A}$ is close to the optimal solution. More specifically, for two solutions we find
\begin{equation}
	| \cO^{(i)}_{A}-\cO^{(i)}_{B}|\lesssim 2\delta \cO^{(i)}\, ,
\end{equation}
while the distance in parameter space can be arbitrarily large. (This is because the mapping from fluxes to fitness is not one-to-one.) However, a notion of convergence in parameter space also turns out to be useful. As explained in \cite{Abel:2014xta}, convergence in parameter space occurs for some variables which form so-called \emph{schemata}, as introduced by Holland \cite{Holland1975}. Although Holland's idea has been criticized over the years\footnote{Based on a more mathematical formulation of GAs, it appears misleading to regard static building blocks as being an intrinsic part of the algorithm's evolution, rather than an artefact of the fitness function. In general, these building blocks change dynamically from one generation to the next, cf. Chs.~3 and 10 of \cite{reeves2002genetic} for a comprehensive discussion. Nonetheless, we generically observe frozen subsets of fluxes characterizing certain regions in moduli space.}, it gives a nice understanding of the generic evolution within a GA, albeit being incomplete. Schemata are characterized by a prototypical behavior of GAs which is that certain parameters quickly converge towards one unifying value. To be more precise, schemata represent the crucial features being favorable characteristics of the optimal solution. This can be interpreted as the fact that a small collection of alleles dominate the fitness. As soon as their values are adjusted correspondingly for the majority of the population, the breeding procedure does not affect them anymore. The formal study of schemata results in the following two conclusions. First, the entire population never quite reaches the optimal solution. This is related to the observation that schemata never touch every single individual{, see Sect.~1.2 in \cite{Abel:2014xta}}. Secondly, mutation plays a crucial role in identifying the best schema. As we will see further below, it forces the algorithm to pick out one schema associated to the ``global'' maximum of $F$. It is important to note that in our case we may not reach the true global maximum, since we typically restrict to certain corners of flux space.

{One string-theoretic example where we might expect the domination of various schema is the toy problem of generating fluxes satisfying the tadpole cancellation condition (\ref{eqn:tadpole}). For large flux integers (relative to $L_{\rm max}$), satisfying (\ref{eqn:tadpole}) requires cancellations between various flux contributions. The various ways flux contributions can cancel represent different (approximate) schema for a search for tadpole-satisfying configurations.}

To study the efficiency of GAs for our problems, we consider the $\ell^1$-distance from an individual to the fittest individual of a generation in flux space,
\begin{equation}\label{eq:HamDist}
D_{A}=\sum_{k=1}^{4(h^{(2,1)}+1)}\, |N_{A}^{k}-N_{\text{ref}}^{k}| ~.
\end{equation} 
Here $\mathbf{N}_{\text{ref}}$ is the flux associated to a generation's fittest individual. Recall that this member is carried over to the next generation due to elitist selection. Thus, $\mathbf{N}_{\text{ref}}$ is only replaced if there appears a member of even greater fitness. In the literature on GAs \cite{jones1995fitness,collard1998fitness}\footnote{See also \cite{Abel:2014xta} for a discussion of FDC in a physics context.}, the utility of a GA for a given problem can be quantified by the \emph{fitness distance correlation} (FDC)\footnote{Notice however that in the present context establishing the FDC in the original sense seems out of reach. This is because $D_{A}$ is in the literature assumed to be the distance \emph{to the nearest global maximum}. To compute $D_{A}$ in this case, we would have to know about the position of \emph{every} solution in advance. If this were the case, we would have started with a deterministic approach right from the beginning. As previously noted, models motivated by string compactifications are most probably NP-hard. We are therefore using a different definition of $D_{A}$ which will allow us to judge whether certain questions are GA-hard or GA-easy.}
\begin{equation}\label{eq:FDC} 
FDC=\dfrac{1}{p}\sum_{A=1}^{p}\, \dfrac{(F_{A}-\bar{F})(D_{A}-\bar{D})}{\sigma_{F}\sigma_{D}}\, .
\end{equation}
Here, $\bar{F}$ is the average fitness, $\bar{D}$ is the average distance, and $\sigma_{F}$, $\sigma_{D}$ are the corresponding standard deviations. One infers that models with
\begin{itemize}
\item $0.15\leq FDC\leq 1$ are GA-hard. This class of models is not tackled well by GAs since the fitness correlates with the distance, i.e., the fitness grows with distance.
\item $-0.15\leq FDC\leq 0.15$ are difficult, that is, there is (almost) no correlation between fitness and distance.
\item $-1\leq FDC\leq -0.15$ are GA-easy. The anti-correlation of fitness and distance implies that the fitness is maximized as the global optimum is approached. Hence, these tasks are suitable for running a GA.
\end{itemize}

When using the algorithm described in Sect.~\ref{sec:GenAlg}, it can be useful to fix certain parameters, while keeping others dynamical. For instance, the width or support of our fitness function can be chosen to decrease to force the population to adapt more and more to a certain value. This is especially helpful when trying to find vacuum solutions sharing certain properties at a given level of accuracy.

To summarize, we can expect the initial population to quickly approach the optimal solution due to the emergence of schemata. Subsequently, there is period of minor adjustments of the remaining flux parameters.

%%%%%%%%%%%%%%%%%%%%%%%%%%%%%%%%%%%%%%%%%%%%%%%
\section{Calabi-Yau hypersurface}\label{sec:Hyper}
%%%%%%%%%%%%%%%%%%%%%%%%%%%%%%%%%%%%%%%%%%%%%%%

In this section, we first review the definition of the hypersurface in question. Next, we explain the general characteristics of our GA. We compare the results to GAs with different breeding mechanisms and to a Metropolis algorithm. We perform only a single search here before we come to more interesting applications in the upcoming section.

\subsection{Expansion around the conifold locus}

In this section we investigate a Calabi-Yau threefold arising as a hypersurface in the weighted projective space ${\bf WP}^{4}_{1,1,1,1,4}$, defined by 
\begin{align}\label{eqn:hyp}
  \sum_{i=1}^4 x_i^8+4x_0^2-8\psi x_0x_1x_2x_3x_4=0\, .
\end{align}
Its Hodge numbers are given by $h^{(1,1)}=1$ and $h^{(2,1)}=149$. We focus on the orientifold taking $x_0\to-x_0,\psi\to-\psi$ along with worldsheet parity reversal, which arises from F-theory compactified on a Calabi-Yau fourfold defined as a hypersurface in ${\bf WP}^5_{1,1,1,1,8,12}$. This amounts to tadpole with $L_{\rm max}=972$ \cite{Giryavets:2003vd}. As described in \cite{Giryavets:2003vd,Giryavets:2004zr,DeWolfe:2004ns}, \eqref{eqn:hyp} has a discrete symmetry group $\Gamma=\mathbb{Z}_8^2\times \mathbb{Z}_2$, and any complex structure deformation except the $\psi$-term is charged under $\Gamma$. By working in a regime where only fluxes consistent with $\Gamma$ are turned on, we can neglect these charged moduli and consistently solve for the periods for the axio-dilaton $\phi$ and uncharged modulus $\psi$.

We are particularly interested in flux vacua near the conifold point $\psi=1$, for which the periods can be written in terms of $x=1-\psi$, $|x|\ll 1$, as \cite{Giryavets:2004zr}
\begin{align}
\cG_{1}(x)&=(2\pi \I)^{3}\left [a_{0}+a_{1}x+\cO(x^{2})\right ]\, ,\nonumber\\
\cG_{2}(x)&=\dfrac{z^{2}(x)}{2\pi\I}\,\ln(x)+(2\pi \I)^{3}\left [b_{0}+b_{1}x+\cO(x^{2})\right ]\, ,\nonumber\\
z^{1}(x)&=(2\pi \I)^{3}\left [c_{0}+c_{1}x+\cO(x^{2})\right ]\, ,\nonumber\\
z^{2}(x)&=(2\pi \I)^{3}\left [d_{0}+d_{1}x+\cO(x^{2})\right ]\, .
\end{align}
Here, the constants $b_{0},d_{0}\in\mathbb{R}$, $a_{0},a_{1},d_{1}\in\I\mathbb{R}$ and $b_{1},c_{0},c_{1}\in\mathbb{C}$ can be found in Sect.~5.1 of \cite{Giryavets:2004zr}. To first order in $x\equiv 1-\psi$, the F-flatness conditions result in
\begin{align}\label{eqn:hyp1}
  \phi&=\frac{f_1\overline{a}_0+f_2\overline{b}_0+f_3\overline{c}_0}{h_1\overline{a}_0+h_2\overline{b}_0+h_3\overline{c}_0}+\mathcal{O}(|x|\ln|x|)\, ,\\
  \ln(x)&=-\frac{2\pi i}{d_1}\left[\frac{(f_1-\phi h_1)(a_1-\frac{\mu_1}{\mu_0}a_0)+(f_2-\phi h_2)(b_1-\frac{\mu_1}{\mu_0}b_0)}{f_2-\phi h_2}+\right.\nonumber\\
  &\left.\quad\quad\quad\quad\quad\quad \frac{(f_3-\phi h_3)(c_1-\frac{\mu_1}{\mu_0}c_0)+(f_4-\phi h_4)d_1}{f_2-\phi h_2}\right]-1\, .\label{eqn:hyp2}
\end{align}
Using Monte Carlo simulations, the authors of \cite{Giryavets:2004zr} explicitly showed that vacua cluster near the conifold point, confirming the expectation from the continuous flux approximation of \cite{Ashok:2003gk}.

In the remainder of this section, we perform all algorithms with
\begin{equation}
\lm=972\kom L=100\, .
\end{equation}
Since we work in the conifold regime, the VEV of the complex structure modulus $x$ is typically driven towards tiny values. This is because the flux vacua cluster around the conifold point.

\subsection{Analyzing different mutation rates}\label{sec:MutHyper}

Let us now find flux vacua with a certain value of the superpotential $\cW_{0}$. This can be relevant when, e.g., looking for vacua that are applicable for any of the three branches of K{\"a}hler moduli stabilisation, namely KKLT \cite{Kachru:2003aw}, LVS \cite{Balasubramanian:2005zx} or K{\"a}hler uplifting \cite{Westphal:2006tn}, which generically require $|\cW_{0}|\ll 1$, $|\cW_{0}|\gtrsim 1$ and $|\cW_{0}|\sim \cO(1\ldots10)$ respectively. For simplicity, we only search for a specific absolute value of $|\cW_{0}|$, although fixing real and imaginary separately is also within the capabilities of our GA. In the following, we use $\cW_{0}$ to denote the absolute value of the superpotential. We run the algorithm with the parameters
\begin{equation} 
\cW_{0}^{\ast}=50000\kom \delta \cW=2000\kom p=1000\kom N_{\text{gen}}=50\, .
\end{equation}
Here, we choose not to apply any crossover, but restrict to the effects of selection and mutation. As we will see in the remainder of this paper, this is already sufficient to find (presumably local) fitness maxima. We discuss the impact of crossover on these results in Sect.~\ref{sec:crossover}. To guarantee the optimal outcome for a GA, it is necessary to adjust the number of mutations as well as the mutation rate. Here, we run our algorithm several times with $n_{\text{mut}}$ mutations per mutated child and mutation rate $q_{\text{mut}}$.

\begin{figure}[t!]
\centering
 \includegraphics[width=0.8\textwidth]{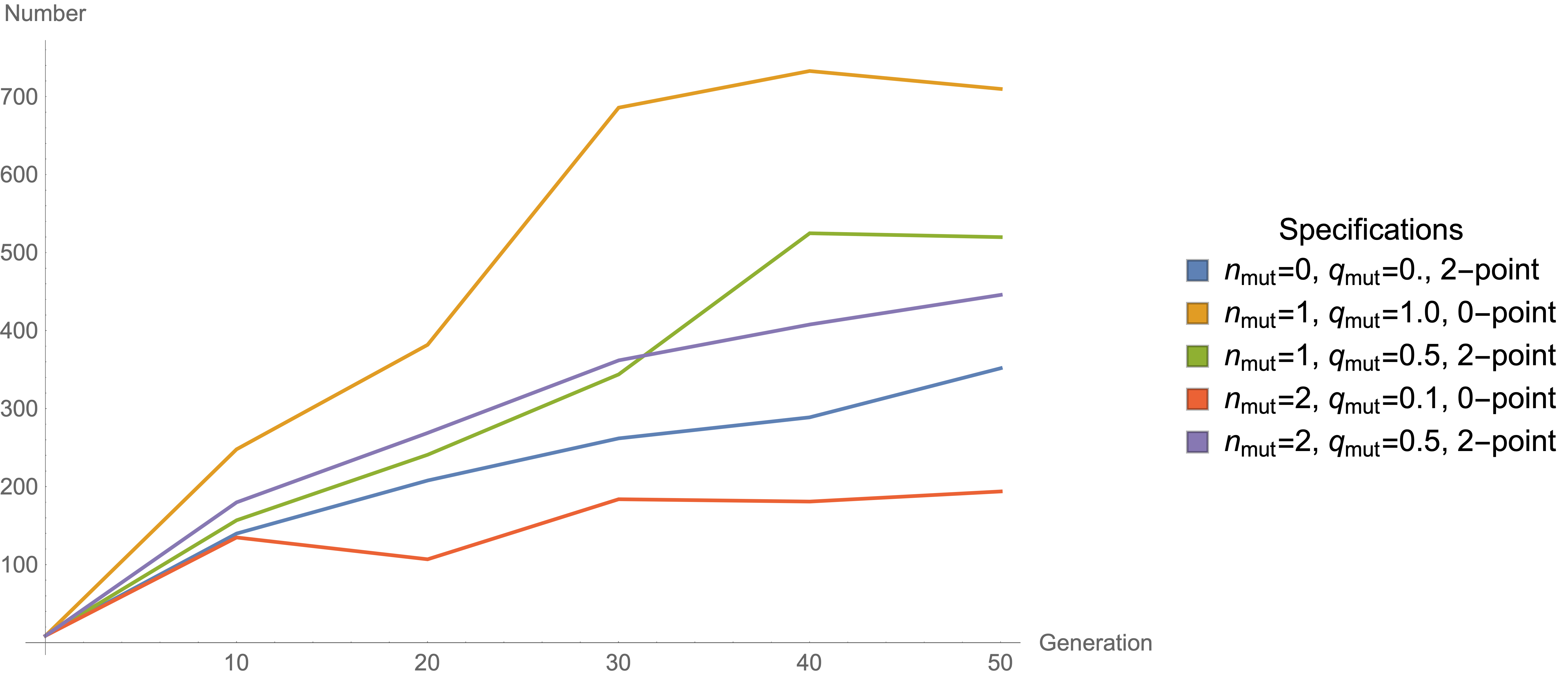}
\caption{Number of members in neighborhoods of radius $r=500$ around $\cW_{0}^{\ast}$ for zero, one or two mutations and different mutation rates.}\label{fig:HMut}
\end{figure}

In order to find the optimal mutation scheme, we compare the convergence around a fixed neighborhood around $\cW_{0}^{\ast}$, cf. Fig.~\ref{fig:HMut}. First of all, we observe that applying two mutations per child (purple line) results in an insufficient convergence rate. This behavior is no surprise because, having only $8$ independent fluxes to begin with, two mutations overshadow the underlying structure of the optimal  solution. In contrast, one mutation results in a seemingly stable evolution over all generations with high convergence rate. Moreover, by comparing the yellow and green line in Fig.~\ref{fig:HMut} we deduce that the search for $\cW_{0}$ considered in this section does not seem to strongly depend on $q_{\text{mut}}$. Nonetheless, we emphasize that it is crucial to scan over different mutation rates in all examples discussed below. Otherwise, the GA's ability to find certain schemata associated to the optimal solution is undermined.

All in all, we conclude that mutation really is an essential ingredient for the success of our algorithm. For our purposes at most one flux mutation per individual seems to be a good guess. However, the mutation rate has to be adjusted accordingly to guarantee the optimal outcome. As we will see below, this typically depends on the specific task.

\subsection{Genetic algorithm dynamics}\label{sec:OverallGA} 

After having investigated different mutation rates, we describe the generic features of a GA's evolution. We proceed by running the algorithm using
\begin{equation}\label{eq:HSimpleWSearch} 
\cW_{0}^{\ast}=50000\kom \delta \cW=2000\kom p=1000\kom N_{\text{gen}}=100\kom q_{\text{mut}}=0.5\, .
\end{equation}
In practice, we are interested in solutions that approximately satisfy $\cW_{0}\approx\cW_{0}^{\ast}$ up to small deviations. We therefore decrease $\delta\cW$ after $50$ generations by $50\%$ every $20$ generations. In doing so, we change the fitness landscape and induce an additional force on the flow of the population. As we will see below, this allows us to find about $93\%$ of the flux vacua localized in the range $[\cW_{0}^{\ast}-50,\cW_{0}^{\ast}+50]$.

\begin{figure}[t!]
\centering
\includegraphics[width=0.8\textwidth]{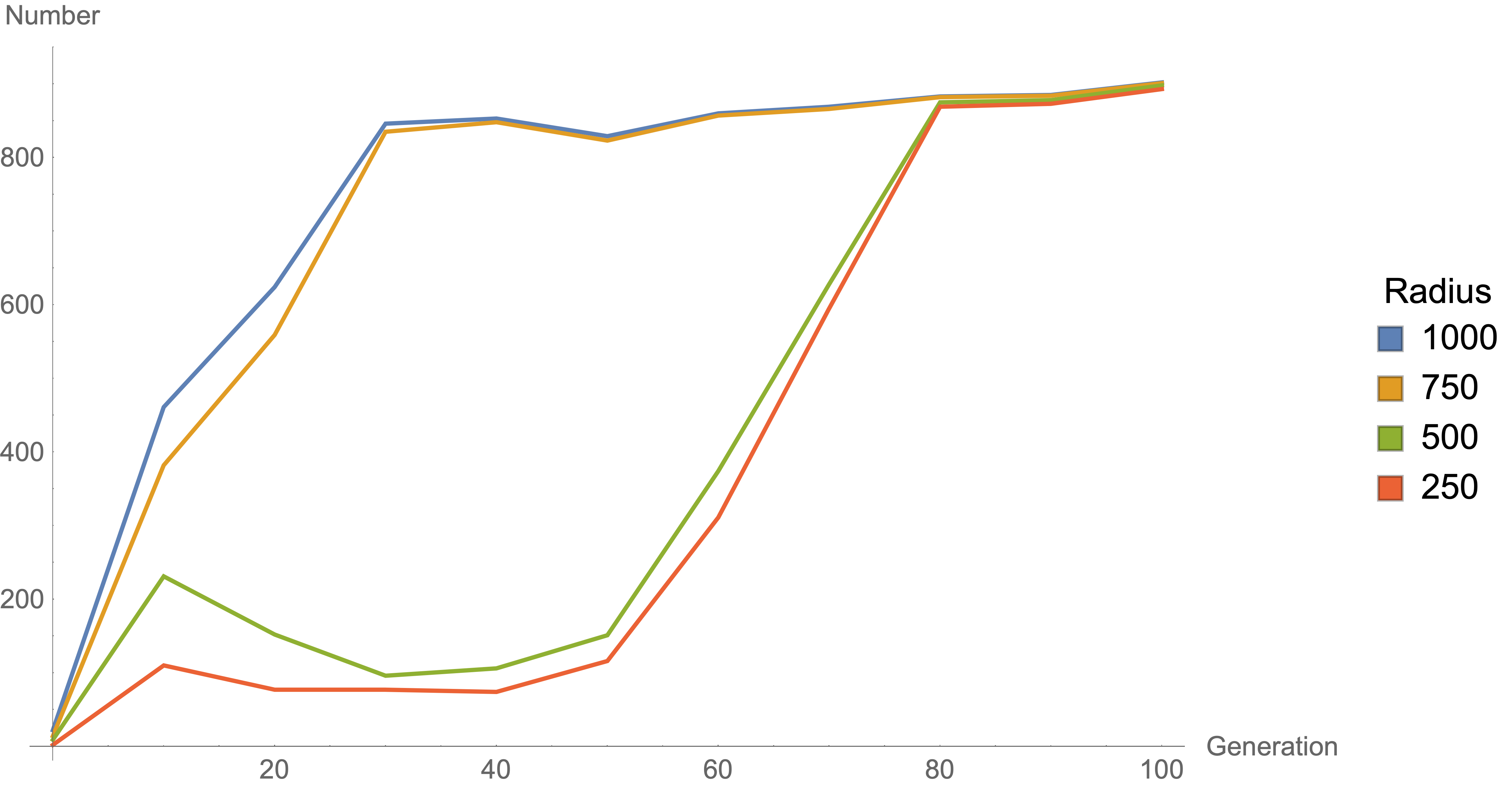}
\caption{Number of individuals within different ranges around $\cW_{0}^{\ast}$.}\label{fig:HP5}
\end{figure}

Fig.~\ref{fig:HP5} shows the number of individuals in neighborhoods of different sizes $r$ around the optimal solution. During the first 50 generations with $\delta\cW=2000$, we observe a sharp incline of the blue and orange curve corresponding to $r=1000$ and $r=750$ respectively. Hence, the population quickly moves towards the optimal solution, but remains mostly outside a neighborhood with $r=500$. After decreasing $\delta\cW$, we observe another sharp incline, but this time in the green ($r=500$) and red ($r=250$) curve. In the final population about $93\%$ of individuals have the optimal value of $\cW_{0}$ to an accuracy of $0.1\%$.

\begin{figure}[t!]
\centering
\includegraphics[width=\textwidth]{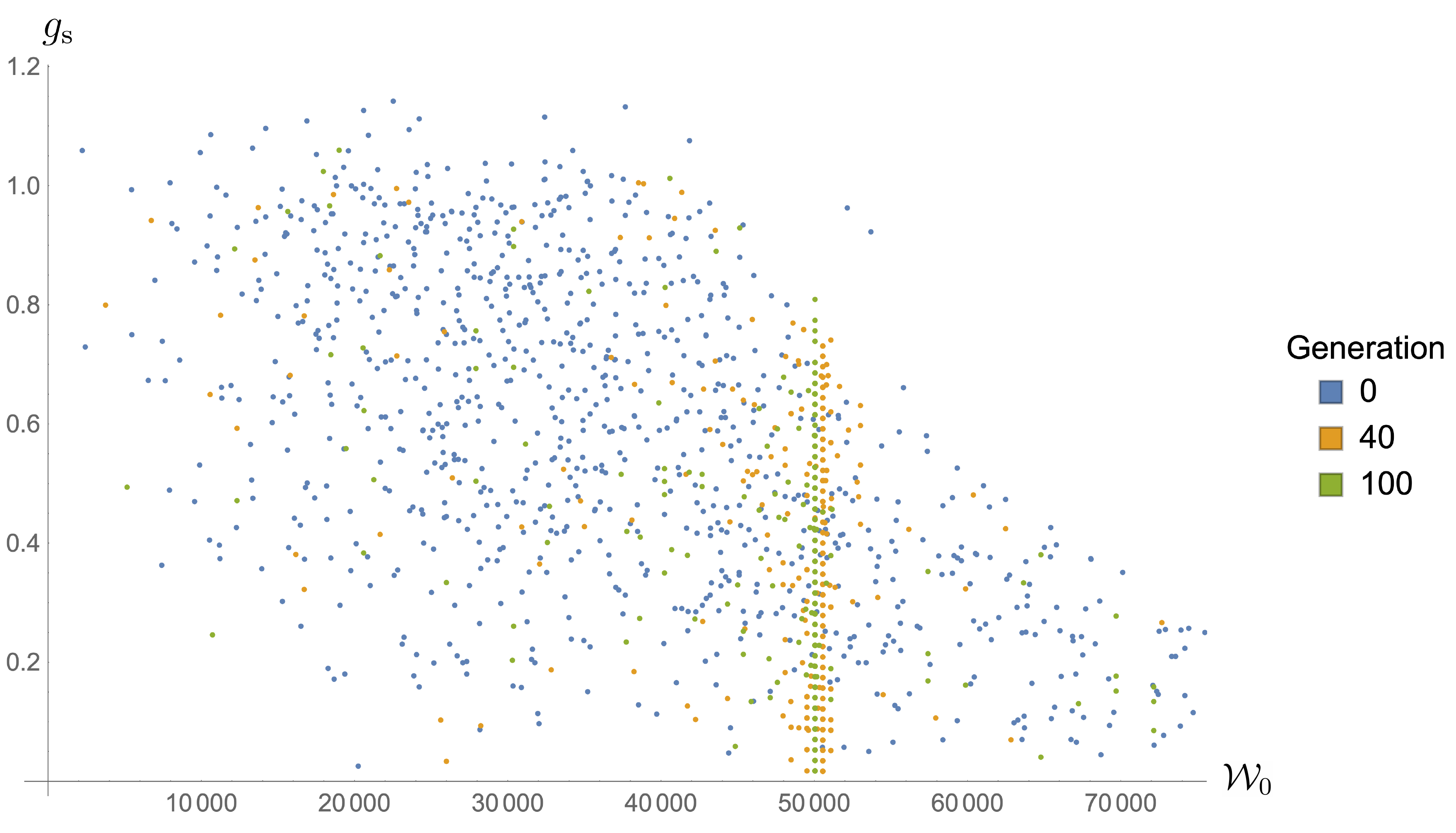}

\vspace*{0.4cm}
\includegraphics[width=\textwidth]{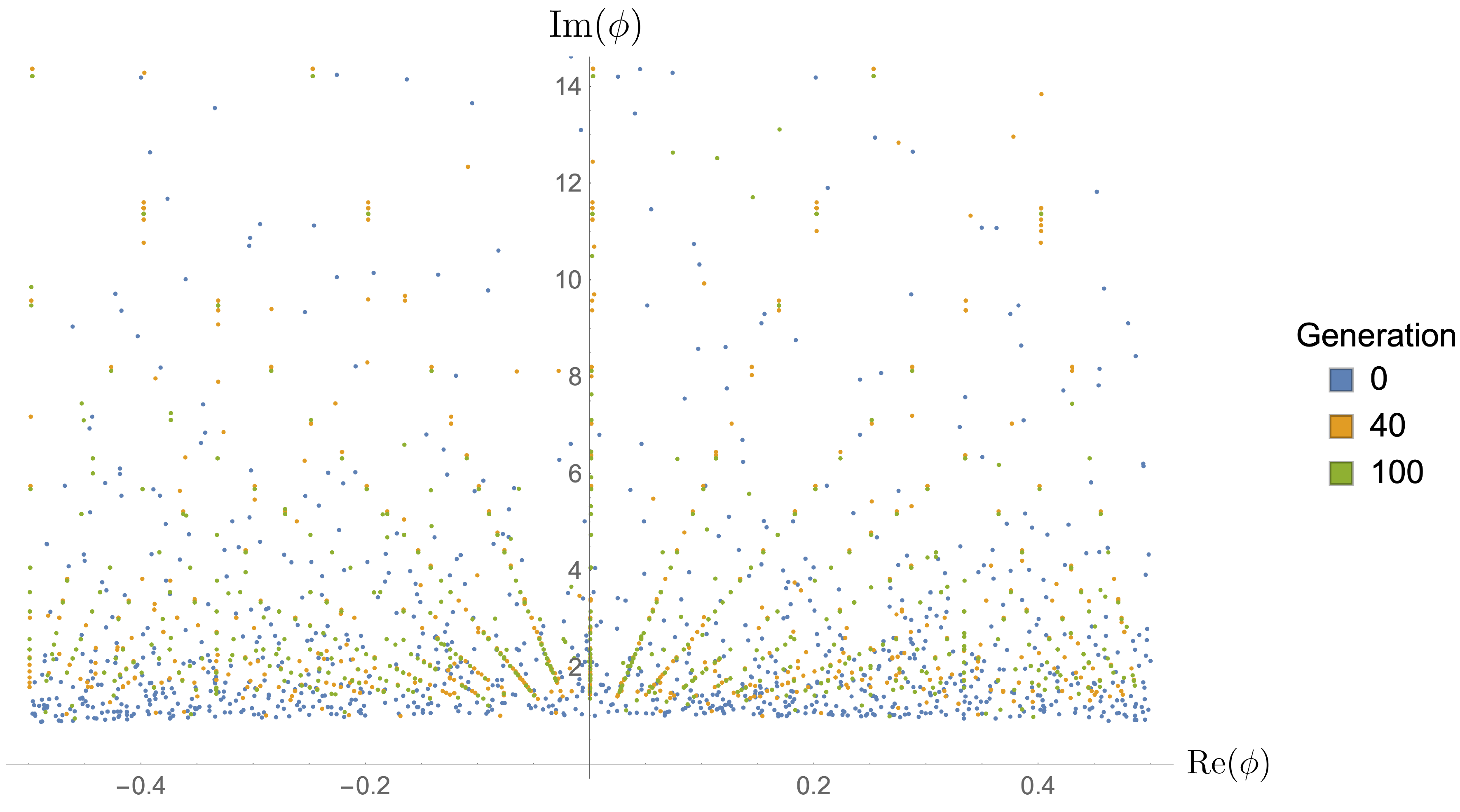}
\caption{Distribution of $\gs$, $\cW_{0}$ and $\phi$ for three generations.}\label{fig:HP3}
\end{figure}

We show the distribution of our population in the $\gs$-$\cW_{0}$-plane on the left of Fig.~\ref{fig:HP3}. At the beginning, the population (blue dots) is randomly scattered across random values of $\cW_{0}$ and $\gs$. After $40$ generations, the distribution of $\cW_{0}$ and $\gs$ exhibits a clear structure in the sense that the individuals around $\cW_{0}^{\ast}$ form almost straight lines in $\gs$-direction. As expected for GAs, a certain fraction of individuals is found far away from the optimal solution. However, the majority of vacua strongly clusters around $\cW_{0}^{\ast}$ which is clearly visible from Fig.~\ref{fig:HP3}. After having decreased $\delta\cW$, the distribution of the final population leaves only one of the previous $\gs$-lines present which is closest to $\cW_{0}^{\ast}$. We interpret this behavior as an indication of improved convergence related to the reduction of $\delta\cW$. The FDC \eqref{eq:FDC} averaged over all generations results in $\text{FDC}=-0.53$ which is within the GA-easy regime. This is in good agreement with the above qualitative observations. Thus, we find reason to believe that our definition of an approximate notion of FDC \eqref{eq:HamDist} is well-suited for our purposes.

Next, let us consider the distribution of $\phi$ in the fundamental domain as depicted on the right hand side of Fig.~\ref{fig:HP3}. Interestingly, after $40$ generations the flux vacua mostly align with straight lines pointing towards the origin. These lines seemingly form a symmetric pattern. On top of that, short vertical lines appear at the outer region of the fundamental domain. This pattern survives until the termination of the algorithm and, even more importantly, is enforced during the evolution. We interpret this result as the manifestation of the schemata associated to the optimal value of $\cW_{0}$. That is, the chromosomes of individual fluxes share specific properties that result in the observed alignment of $\phi$-VEVs. This interpretation is further supported by comparing different crossover and breeding mechanisms in the next two sections.

\begin{figure}[t!]
\centering
 \includegraphics[width=0.5\textwidth]{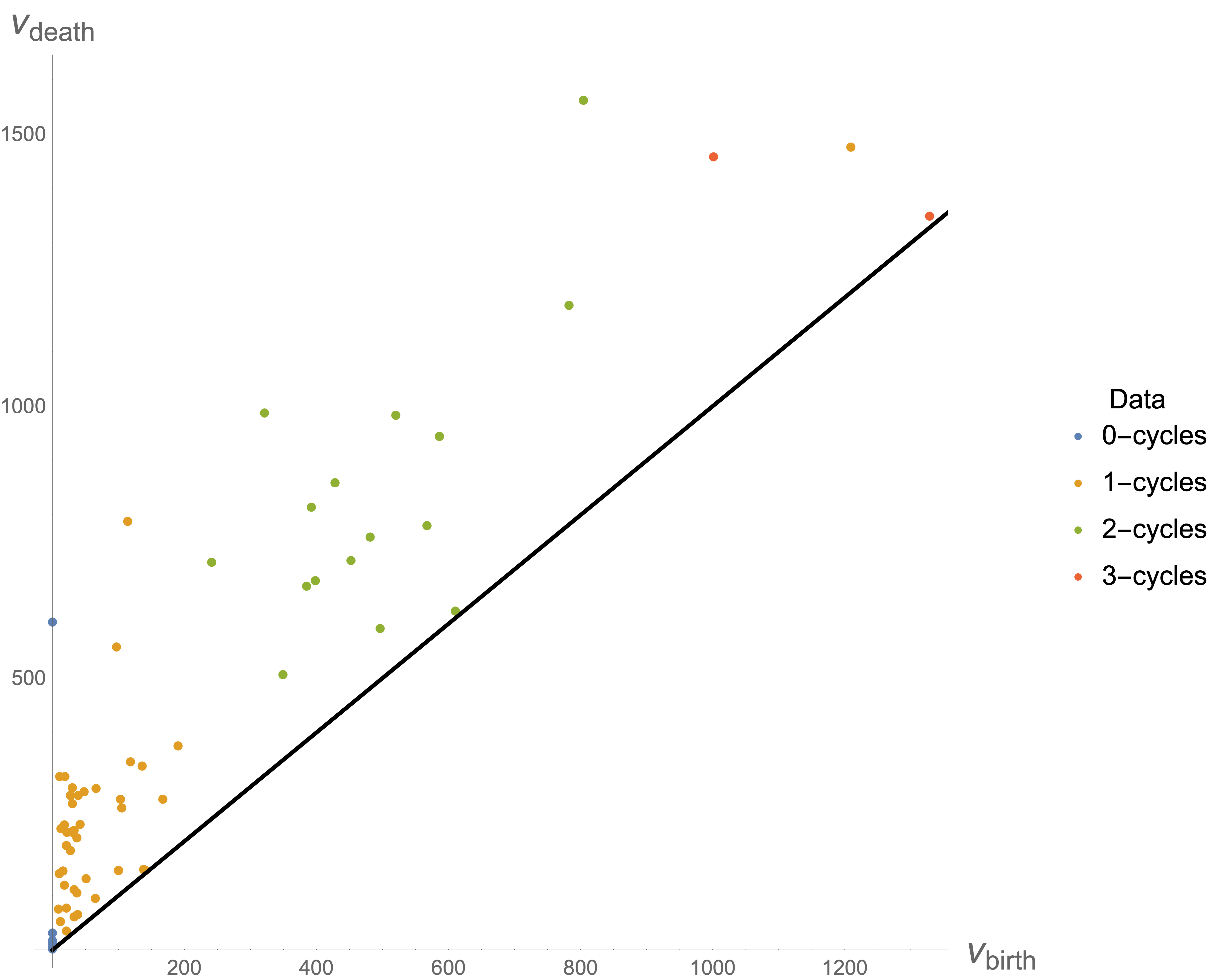}\includegraphics[width=0.5\textwidth]{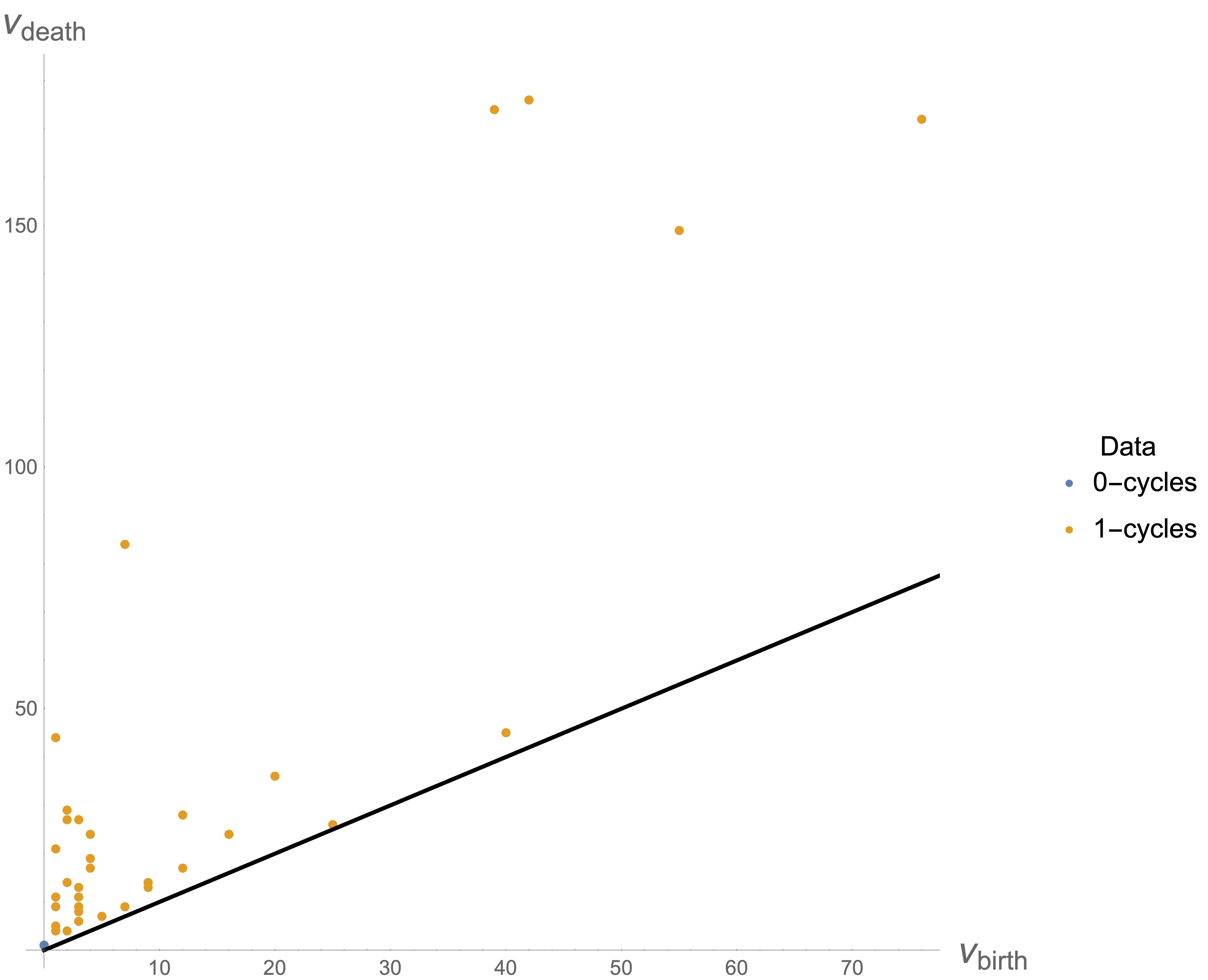}
\caption{Persistence diagram for the population in the initial (left) and final (right) generation in flux space.}\label{fig:HP1}
\end{figure}

To understand these results further, we use \emph{persistent homology} to analyze the topological structure of the distribution in flux space, applying the methods used in \cite{Cole:2018emh,Cirafici:2015pky}, see also \cite{Edelsbrunner2002,zomorodian2005computing,carlsson2009topology,carlsson_2014,Cole:2017kve,Murugan:2019alb} for more elaborate introductions. Persistent homology has previously been used to characterize distributions of string vacua, and identify loci in moduli space related to phenomenologically interesting properties. 

Applied to the scenario under consideration, we compare the flux distributions of populations at different generations during the GA's evolution. We specifically consider the persistence diagrams associated to the data set of fluxes for a single generation. These diagrams encode the two scales at which certain higher-dimensional topological features in the set of fluxes first emerge and vanish. In Fig.~\ref{fig:HP1}, we compare the flux configurations of the initial (top) and the final (bottom) population.\footnote{More specifically, we consider the distribution of \emph{constrained} fluxes, cf. Sect.~\ref{sec:Binary} for details. These fluxes are eventually fixed by evolution of the GA. Restricting to the constrained fluxes allows one to observe the emergence of schemata during the GA's evolution.} The plot on the top shows a lot of structure in terms of the presence of various $k$-cycles, albeit most die very quickly. Randomly choosing the initial fluxes results in regions of non-trivial topology. In contrast, the distribution of the final population on the bottom of Fig.~\ref{fig:HP1} does not show much topologically interesting structure. Moreover, the majority of observed features are very short-lived in terms of the associated scales. This indicates that (most) fluxes form a cluster which does not exhibit any non-trivial topological properties at large scales.

\subsection{The role of crossover}\label{sec:crossover}

\begin{figure}[t!]
\centering
\includegraphics[width=0.8\textwidth]{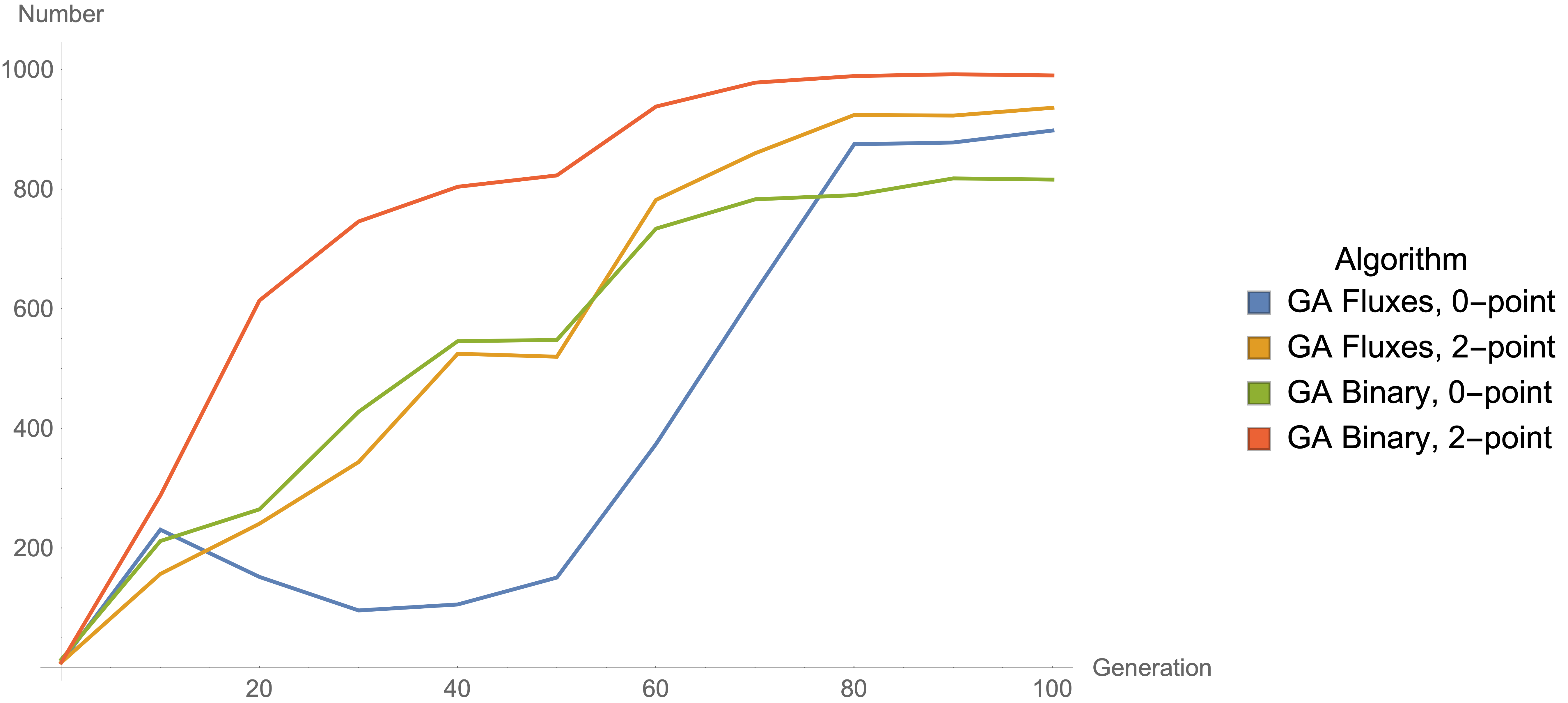}
\caption{The number of individuals within a range of $500$ around $\cW_{0}^{\ast}$ for the algorithms of Sects.~\ref{sec:OverallGA}, \ref{sec:crossover} and \ref{sec:Binary}.}\label{fig:HP2}
\end{figure}

The takeaway message of the previous section is that a GA can be applied even without performing any crossover. The true power of GAs typically comes only to light when having a dynamical interaction of selection, crossover and mutation. In the remainder of this section, we would like to contrast different choices for running the GA. For instance, one can consider various crossover operators at the same time to increase the GA's performance. This is necessary since different tasks generally require the application of suitable crossover procedures. Here, we restrict to a $2$-point crossover operator as explained in Sect.~\ref{sec:GenAlg}.

Using the same initial population as previously, we find the evolution depicted as the orange line in Fig.~\ref{fig:HP2}. If we compare the blue (no crossover) and orange ($2$-point crossover) line, we observe that applying a non-trivial crossover operator leads to a more stable evolution, especially at early stages. Recall that we reduce the support of the fitness at generation $50$. At this point of the GA's evolution without crossover the population seemingly feels a strong pull towards the optimal solution. In contrast, the convergence rate with $2$-point crossover increases almost monotonically. Thus, although the final population is qualitatively the same, we find significant difference in the overall evolution.

The previously mentioned structures in Fig.~\ref{fig:HP3} emerge with $2$-point crossover as well which is a strong indication for a characteristic feature associated to the optimal solution, see the discussion in the upcoming section. We also analyzed the resulting distribution of constrained fluxes in the spirit of the previous section. Indeed, the persistence diagram for the final population mirrors the results of Fig.~\ref{fig:HP1}.

\subsection{Comparing different breeding mechanisms}\label{sec:Binary} 

There is another way of implementing a genetic algorithm which might be useful when large input parameters appear. Recall that, in order to guarantee that the algorithm can access most of the landscape, we have to ensure that the whole range of alleles appears within each individual locus of a chromosome. This means that the necessary population size scales with the maximally allowed allele. Hence, it might be useful to interpret each flux as a single chromosome by encoding it in a binary form. As opposed to the breeding and crossover mechanism explained in Sect.~\ref{sec:GenAlg}, we now perform crossovers for \emph{each individual} flux. That is, we act on $8$ strings of zeros and ones. It is expected that the necessary population size is drastically reduced in this way. In fact, one finds that for a chromosome of length $\ell$ \cite{Abel:2014xta}
\begin{equation}\label{eq:PBinary} 
p_{\rm min}\approx 1+\dfrac{1}{\log(2)}\, \log\left (\dfrac{-\ell}{\log(P_{*})}\right )
\end{equation}
with $P_{*}$ being the probability with which every allele should appear at each locus at least once. This is generally much smaller than the minimum population size for alleles taking values in $[-L_{\rm max},L_{\rm max}]$.

In this section, we compare the performance of both methods, applying $0$- and $2$-point crossover for the same population size $p=1000$.\footnote{The precise value of a minimal population size does not matter in our simple models. However, for applications with more degrees of freedom the advantage of a reduced minimal $p$ from binary encoding can be substantial. For our purposes, this way of encoding crossovers gives similar results.}
First, the overall evolution is very alike to the distribution shown in Fig.~\ref{fig:HP3}. It is worth mentioning that the observed pattern in the distribution of $\phi$-VEVs also emerges here independently of whether we perform $0$- or $2$-point crossover. Thus, we are convinced that this linear structure is not related to the applied crossover procedure, but rather a property of the vacua close to the optimal solution.

In Fig.~\ref{fig:HP2}, we see that there are only minor differences in the overall performance. Hence, this way of performing crossovers and mutations might be advantageous in higher-dimensional models than considered here. The point is that solving the equations of motion in high-dimensional spaces is computationally very costly. Thus, working with smaller population sizes provides a welcome reduction in computational expense.

As before, we observe a strong clustering also for the case of binary crossover. This becomes apparent by performing a \emph{Principal Component Analysis} (PCA) on the final population. In general, PCA is useful for determining sets of linearly uncorrelated variables. Applied to the flux distribution of our population we gain information about the effective dimensionality of the resulting (discrete) hypersurface. That is, some fluxes will take specific values as we argued previously in the context of schemata in Sect.~\ref{sec:ConvEv}, while others remain arbitrarily distributed over some range of flux values $[-L,L]$. {PCA organizes the full flux space into orthogonal dimensions of decreasing variance.} 
%The output of the PCA is a set of numbers quantifying how many individuals (in $\%$) can be found along each principal component. 
{The intuition is that most data actually lives on a hyperplane of smaller dimension than the naive dimensionality of the data set. This can be quantified by examining the percentage of the data's variance that is retained when the data is projected onto a particular hyperplane.}
%The latter are typically linear combinations of fluxes, i.e., the numbers displayed below are not directly linked to original flux basis.

{One subtlety is that our gauge-fixing condition forces different flux components to take values in ranges of different scales. Naive application of PCA would then artificially inflate the importance of fluxes taking values in a large range. To remove this artifact, we first rescale the flux components in the initial population so that each component has zero mean and unit variance. The same scaling transformation is then applied to the final transformation.}

For the final population obtained in Sect.~\ref{sec:OverallGA} and Sect.~\ref{sec:crossover}, we find that the fluxes live on an effectively $1$-dimensional hypersurface, with 92.84\% (no crossover) and 89.57\% (2-point crossover) of the variance in the first component. Similarly, binary encoding leads to significantly more clustering with 99.63\% {(no crossover) and 92.96\% (2-point crossover)} variance in the first component. Qualitatively, both crossover procedures give rise to a very similar evolution of the variance.

\subsection{Comparison to a Metropolis algorithm}\label{sec:Metro} 

In this section, we compare the results of our genetic algorithm to those of random walk approaches. More specifically, we consider a \emph{Metropolis algorithm} \cite{doi:10.1063/1.1699114} or rather its related variant of \emph{simulated annealing} \cite{kirkpatrick1984simulated}. As before, we start with a randomly sampled generation of flux vacua. In contrast to the GA, we take a random step from each point in flux space until we find a new solution to the F-term equations with small enough tadpole. Explicitly, if $\mathbf{N}\in\mathbb{Z}^{8}$ is some chromosome in our population, then a new flux $\mathbf{M}\in\mathbb{Z}^{8}$ is chosen as
\begin{equation}
\mathbf{M}=\mathbf{N}+r\mathbf{q}
\end{equation}
where $\mathbf{q}\in\mathbb{Z}^{8}$, $|q_{i}|\leq 1$, is a random vector and $r\in\mathbb{N}$ some fixed step size. The vector $\mathbf{q}$ must be chosen such that $\mathbf{M}$ also leads to a solution of the F-term constraints with $N_{\text{flux}}(\mathbf{M})\leq \lm$. Practically, this means that we have to perform many such steps and test the criteria for a viable physical flux choice. 

In order to define a new population, we evaluate an energy functional $E$ given by the difference in fitness
\begin{equation}
E=F(\mathbf{N})-F(\mathbf{M})\, .
\end{equation}
If $E\leq 0$, the new configuration $\mathbf{M}$ is energetically favorable. In this case, $\mathbf{M}$ is carried over to the new population. If $E>0$, we define a probability
\begin{equation}
P_{T}(E)=\mathrm{e}^{-E/T}
\end{equation}
with some ``temperature'' $T$ and draw a random number $q\in [0,1]$. If $q<P_{T}(E)$, $\mathbf{M}$ is chosen as a member of the next generation.

\begin{figure}[t!]
\centering
\includegraphics[width=\textwidth]{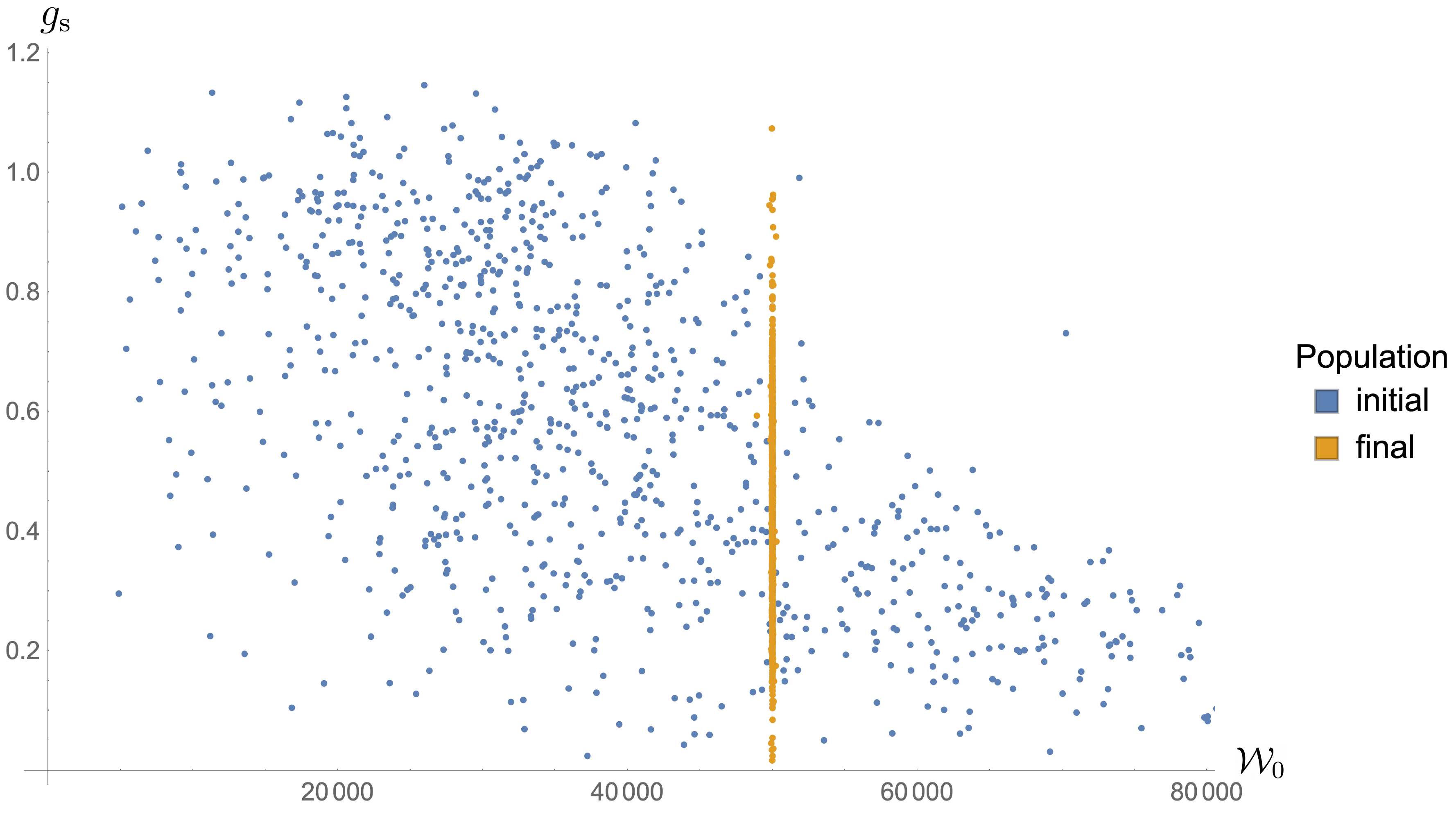}
\caption{The distribution for $\gs$ and $\cW_{0}$ obtained from a Metropolis algorithm.}\label{fig:HP4}
\end{figure}

Again, we use the same initial flux choices and let the algorithm run for $1000$ steps for each individual. We choose $r=1$ and {determine the initial temperature $T$ for each vacuum solution individually. That is, we fix an initial acceptance rate of about $20\%$ by considering $100$ random steps $\mathbf{q}$ before starting the evolution. We decrease the temperature by a factor of $10$ every $100$ steps. In total, we find that $96\%$ of individuals can be found within the range $[\cW_{0}^{\ast}-50,\cW_{0}^{\ast}+50]$.} If we compare the overall evolution of the distribution of $\gs$ and $\cW_{0}$ in Fig.~\ref{fig:HP4} with the one for the GA in Fig.~\ref{fig:HP3}, it becomes clear that this method, with each individual walking independently of the others, {does not exploit correlations between different vacua. Although the population gets close to the optimal solution, it shows less structure than the equivalent plots in Fig.~\ref{fig:HP3}. Similarly, we checked that} the distribution of $\phi$-VEVs does not exhibit the symmetric pattern of Fig.~\ref{fig:HP3}. Qualitatively, this can be understood as follows. After some generations within a GA, all vacua are in a sense related to each other due to the applied selection process. On the contrary, the random walk approach does not link solutions in a similar fashion. Thus, there is no reason for them to share characteristics, albeit the population gathers around the optimal value $\cW_{0}^{\ast}$.

\begin{figure}[t!]
\centering
\includegraphics[width=0.5\textwidth]{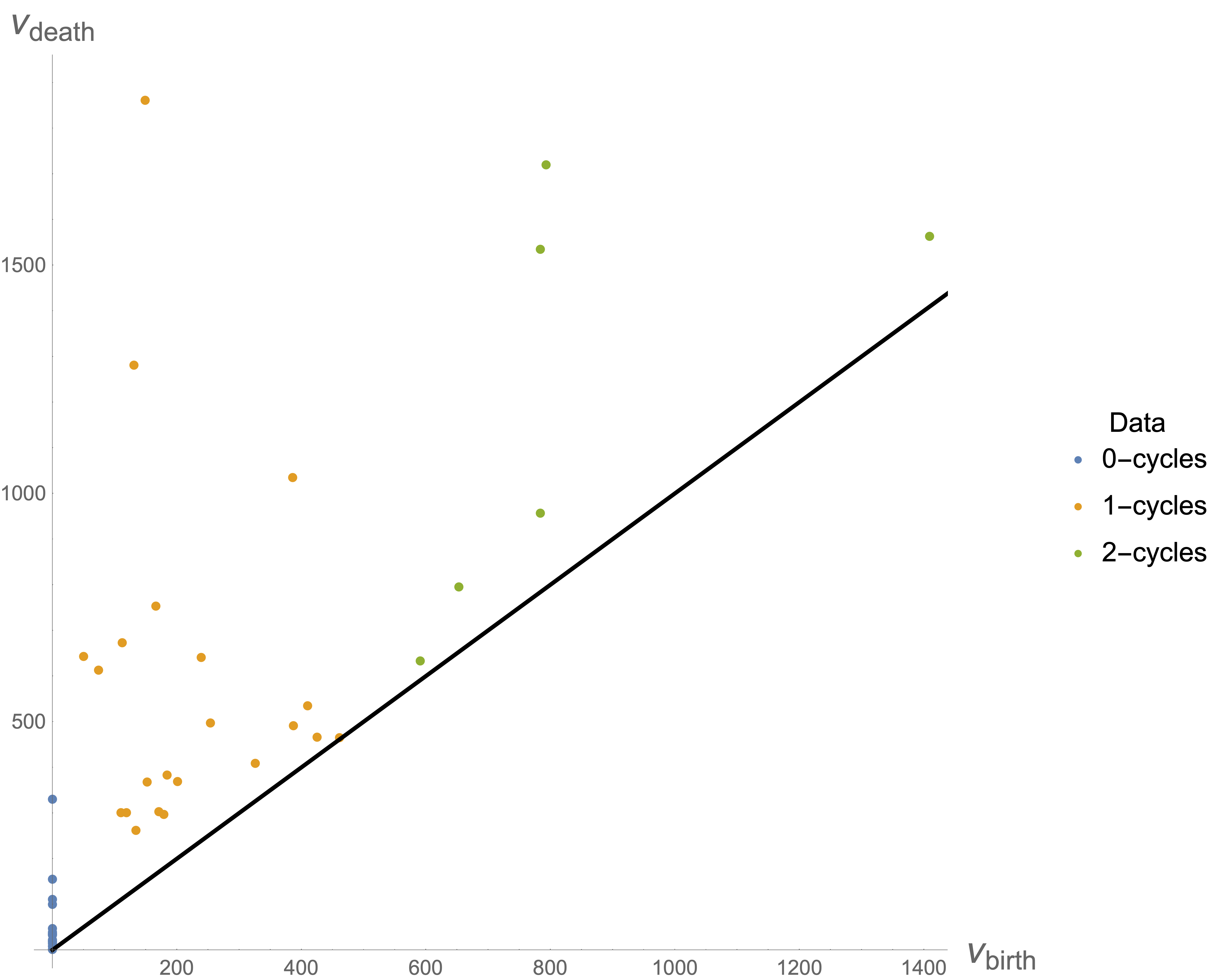}
\caption{The persistence diagram for the final population obtained with the Metropolis algorithm. There is significantly more large scale structure to be observed than in the analogous plot on the bottom of Fig.~\ref{fig:HP1}.}\label{fig:HPMet1}
\end{figure}

Since the final population is randomly distributed around the optimal solution, there should be less clustering in flux space. By this reasoning, we should expect the distribution of fluxes in the final distribution to have a much richer structure than for a GA. This can be seen by comparing the persistence diagrams. Recall that in Fig.~\ref{fig:HP1} we observed that the final population resembles a cluster in flux space. In contrast, the persistence diagram of the final population associated to the Metropolis algorithm in Fig.~\ref{fig:HPMet1} exhibits more long-lived higher-dimensional features. That is, there are a greater number of long-lived $1$- and $2$-cycles.

{As before, we performed PCA on the flux distribution. Here, the variance { in the first component} increases to only {$54.6\%$} in contrast to {$\gtrsim 89\%$} for the previous algorithms. That is, the final population obtained via the Metropolis algorithm is less constrained in flux space.} {This is to be expected from dynamics of the algorithm and already confirmed via TDA.}
%Note that {the appearance of a dominating component within the PCA} is related to our gauge choice. 

{Finally, we note that in our experiments, Metropolis was less efficient than GAs in two senses. First, the random steps taken by Metropolis were less efficient in finding physical vacua than our GA. As previously argued, crossover should be more efficient in finding physical vacua than taking random steps. Second, even in terms of physical steps, Metropolis took longer than our GAs. We found that it took Metropolis an average of 336 physical steps to match the performance of a GA after 80 generations. }

%%%%%%%%%%%%%%%%%%%%%%%%%%%%%%%%%%%%%%%%%%%%%%%
\section{Symmetric $T^{6}$}\label{sec:T6}
%%%%%%%%%%%%%%%%%%%%%%%%%%%%%%%%%%%%%%%%%%%%%%%

In this section we consider compactifications on a symmetric $T^6$. One interesting feature of the symmetric $T^6$ is the existence of vacua with vanishing tree-level superpotential. These special vacua give rise to the opportunity to study their emergence within GAs. A major issue with finding such vacua is that they are quite scarce in comparison to generic vacua. Hence, we cannot take for granted that GAs can identify these exceptional cases within the landscape.

\subsection{Generic and special vacua}

We follow the conventions of \cite{DeWolfe:2004ns}. The symmetric torus can be viewed as a direct product of three $T^{2}$ setting the modular parameters $\tau\equiv \tau_{1}=\tau_{2}=\tau_{3}$ all equal. The moduli space has two complex dimensions, so we have upon gauge fixing $8$ independent flux parameters. Let us first parametrize a general $T^6$ before we specialize to the symmetric case further below. We define coordinates $x^i,y^i$ for $i=1,2,3$ with periodicity $x^i\equiv x^i+1,y^i\equiv y^i+1$ such that the three holomorphic 1-forms can be written as $\dif z^i=\dif x^i+\tau^{ij}\dif y^j$. We take the orientation
\begin{align}
	\int \dif x^1\wedge \dif x^2\wedge \dif x^3\wedge \dif y^1\wedge \dif y^2\wedge \dif y^3=1
\end{align}
and choose a symplectic basis for $H^3(T^6,\mathbb{Z})$, namely
\begin{align}\label{eq:SympBasis} 
	\alpha^0&=\dif x^1\wedge \dif x^2\wedge \dif x^3\kom &\alpha_{ij}&=\frac{1}{2}\epsilon_{ilm}\dif x^l\wedge \dif x^m\wedge \dif y^j\, ,\nonumber\\
	\beta^{ij}&=-\frac{1}{2}\epsilon_{jlm}\dif y^l\wedge \dif y^m\wedge \dif x^i\kom 	&\beta^0&=\dif y^1\wedge \dif y^2\wedge \dif y^3\, .
\end{align}
The holomorphic 3-form can be written as
\begin{equation}
\Omega=\dif z^1\wedge \dif z^2\wedge \dif z^3\, .
\end{equation}
We can expand the $3$-form fluxes in terms of the symplectic basis \eqref{eq:SympBasis}
\begin{align}
	F_3&=a^0\alpha^0+a^{ij}\alpha_{ij}+b_{ij}\beta^{ij}+b_0\beta^0\nonumber\\
	H_3&=c^0\alpha^0+c^{ij}\alpha_{ij}+d_{ij}\beta^{ij}+d_0\beta^0\, .
\end{align}

For a symmetric $T^{6}$, we take
\begin{equation}
\tau^{ij}=\tau\delta^{ij}\, .
\end{equation}
This is equivalent to taking the $T^6$ to be factorizable as three two-tori with equal modular parameter. Similarly, the fluxes get reduced to
\begin{align}\label{eq:FluxDefT6} 
	a^{ij}=a\delta^{ij},\quad b_{ij}=b\delta_{ij},\quad c^{ij}=c\delta^{ij},\quad d_{ij}=d\delta_{ij}\, .
\end{align}
The superpotential takes the simple form
\begin{align}
	\cW=P_1(\tau)-\phi P_2(\tau)
\end{align}
where $P_i$ are cubic polynomials in $\tau$, i.e.,
\begin{align}
	P_1(\tau)&= a^0\tau^3-3a\tau^2-3b\tau-b_0\, ,\\
	P_2(\tau)&= c^0\tau^3-3c\tau^2-3d\tau-d_0\, .
\end{align}
The K\"ahler potential for $\tau$ and $\phi$ reads
\begin{align}
	\mathcal{K}=-3\log(-i(\tau-\overline{\tau}))-\log(-i(\phi-\overline{\phi}))
\end{align}
and the D3-brane charge induced by fluxes corresponds to
\begin{align}
	N_{\rm flux}=b_0c^0-a^0d_0+3(bc-ad)\, .
\end{align}

The F-term constraints can be written in the form
\begin{align}
	\label{eq:FTerm1} P_1(\tau)-\overline{\phi}P_2(\tau)&=0\, ,\\
	\label{eq:FTerm2} P_1(\tau)-\phi P_2(\tau)&=(\tau-\overline{\tau})(P_{1}^{\prime}(\tau)-\phi P_{2}^{\prime}(\tau))
\end{align}
For non-zero VEV of the superpotential $\cW_{0}\neq 0$, the axio-dilaton can be obtained using \eqref{eq:FTerm1} so that
\begin{equation}
\phi=\dfrac{\overline{P_{1}(\tau)}}{\overline{P_{2}(\tau)}}\, .
\end{equation}
Plugged into \eqref{eq:FTerm2}, one finds for $\tau=x+iy$
\begin{align}
  q_1(x)y^2&=q_3(x)\label{eqn:t6y1}\, ,\\
  q_0(x)y^4&=q_4(x)\label{eqn:t6y2}\, .
\end{align}
The $q_i$ are polynomials in $x$ which have for instance been computed in the appendix of \cite{DeWolfe:2004ns}. Surprisingly, multiplying both \eqref{eqn:t6y1} and \eqref{eqn:t6y2} to eliminate $y$, a cubic (rather than sextic) equation in $x$ remains so that $x$ can be found by solving
\begin{align}
\alpha_3 x^3+\alpha_2 x^2+\alpha_1 x+\alpha_0=0\label{eqn:t6x}\, .
\end{align}
The coefficients $\alpha_i$ are combinations of flux integers and can again be found in \cite{DeWolfe:2004ns}.

Solutions with $\cW_{0}=0$ satisfy
\begin{equation}
P_{1}(\tau)=P_{2}(\tau)=0\, .
\end{equation}
Thus, the solution for $\phi$ simply reads
\begin{equation}
\phi=\dfrac{P_{1}^{\prime}(\tau)}{P_{2}^{\prime}(\tau)}\, .
\end{equation}
As shown in \cite{Kachru:2002he}, these solutions obey the special property that $P_{1}$ and $P_{2}$ must factorize over the integers, cf. Sect.~4.3.3 in \cite{DeWolfe:2004ns}. Recently, it was shown in \cite{Cole:2018emh} using persistent homology that the solutions with $\cW_{0}=0$, when combined with flux quantization and tadpole cancellation, exhibit a different structure in the moduli space than generic vacua. We will see in Sect.~\ref{sec:T6W0} that this also plays an important role when applying GAs to minimizing $\cW_{0}$.

\subsection{Searching for $g_{\text{s}}$ -- Emergence of multiple schemata}\label{sec:T6G} 

In this section, we discuss the emergence of various correlations and schemata within the GA's evolution. For the sake of simplicity, let us first look for solutions with a certain value of $g_{\text{s}}$. We begin with the following choice of parameters
\begin{equation}
g_{\text{s}}^{\ast}=0.3\kom \delta g_{\text{s}}=0.05\kom p=10000\kom N_{\text{gen}}=100\kom \lm=500\kom q_{\text{mut}}=1\, .
\end{equation}
We use a Gaussian to define the fitness, but e.g. a combination of Heaviside-functions works equally well. After half of the generations, we divide $\delta \gs$ by a factor of $2$ every $10$ generations. We employ no crossover within this search since it appears to be more efficient than employing a $2$-point crossover operator.

\begin{figure}[t!]
\centering
 \includegraphics[width=\textwidth]{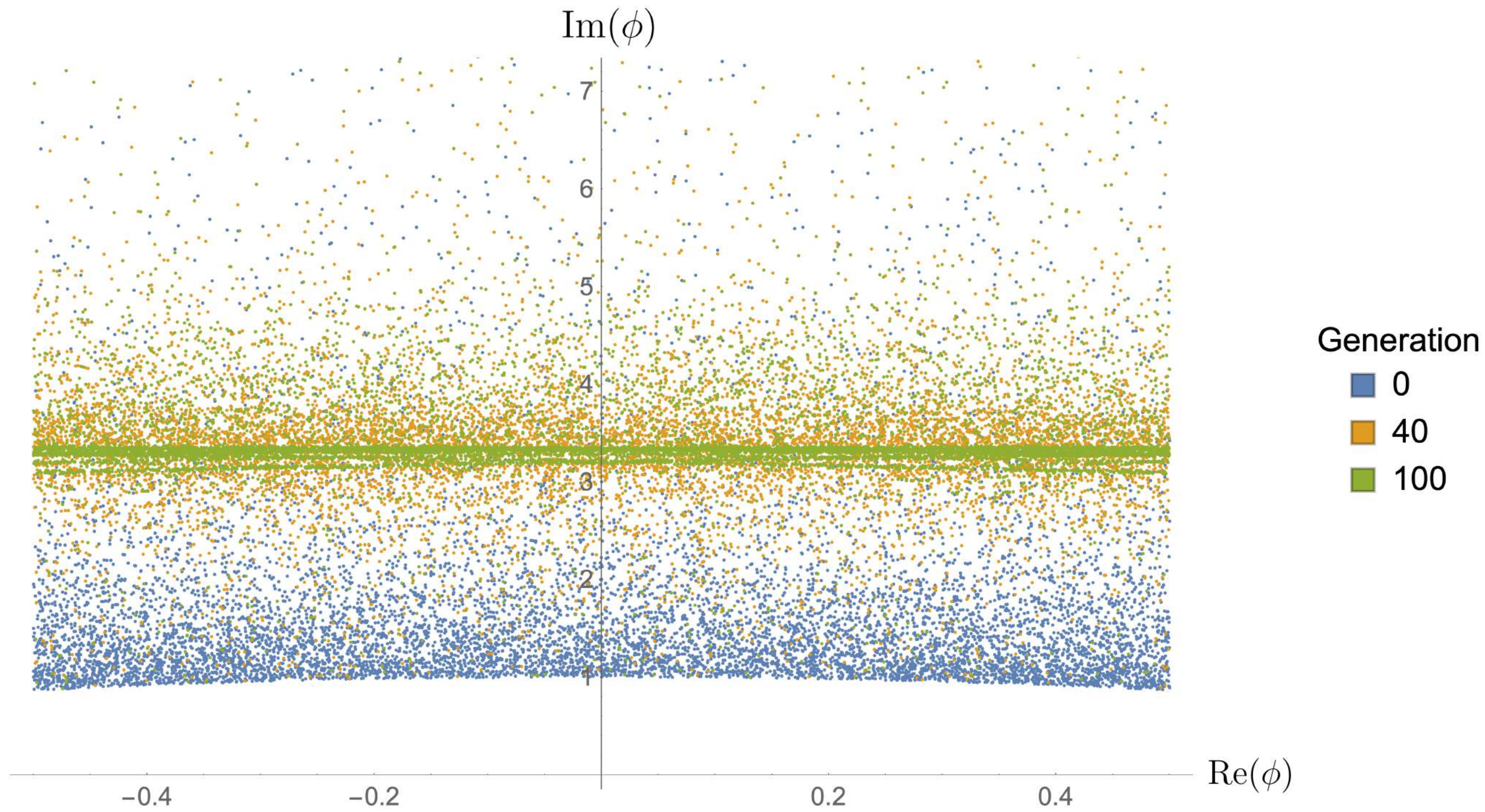}
 
 \vspace*{0.4cm}
 \includegraphics[width=\textwidth]{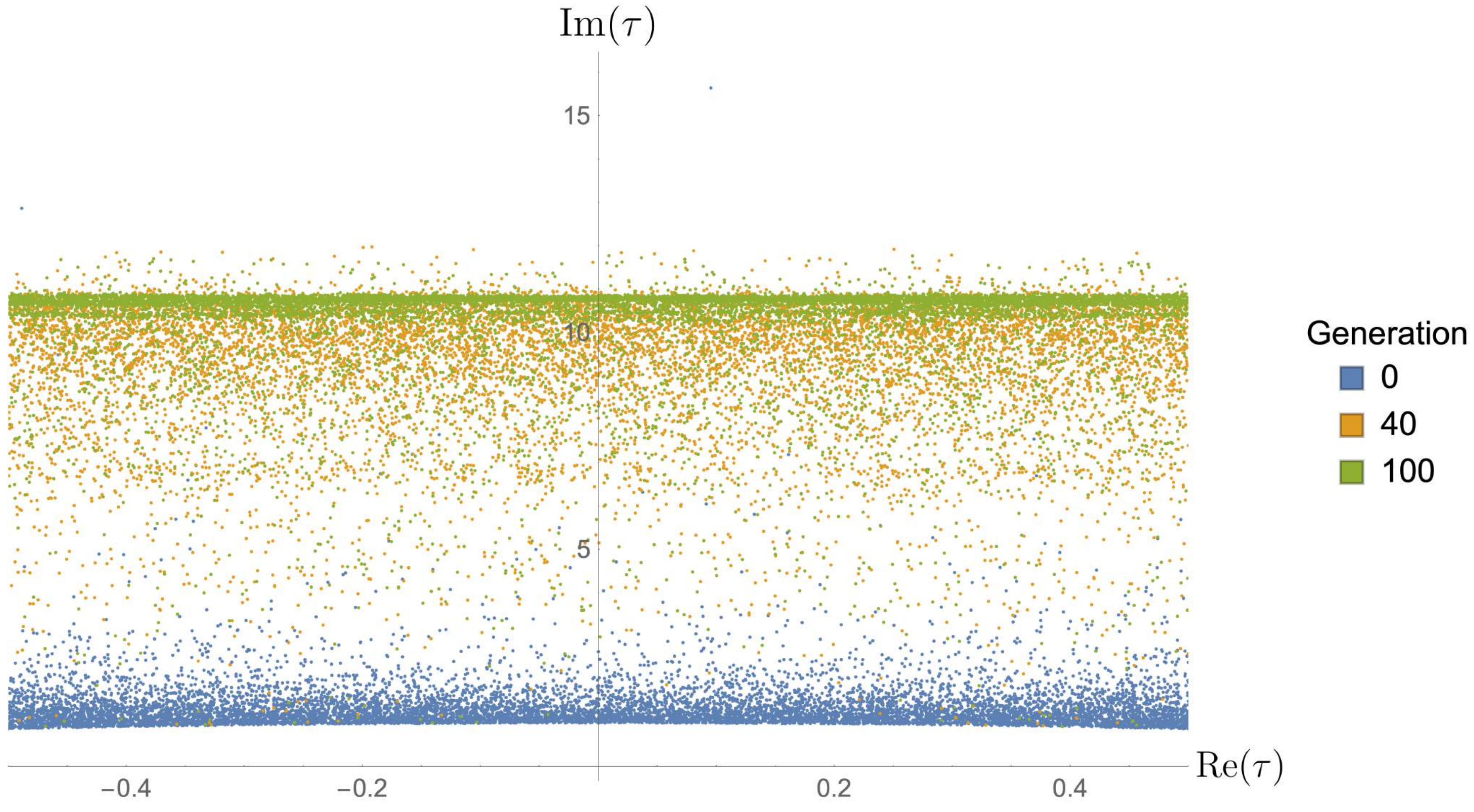}
\caption{Evolution of the distribution of $\phi$ (top) and $\tau$ (bottom) for three generations.}\label{fig:1}
\end{figure}

\begin{figure}[t!]
\centering
 \includegraphics[width=\textwidth]{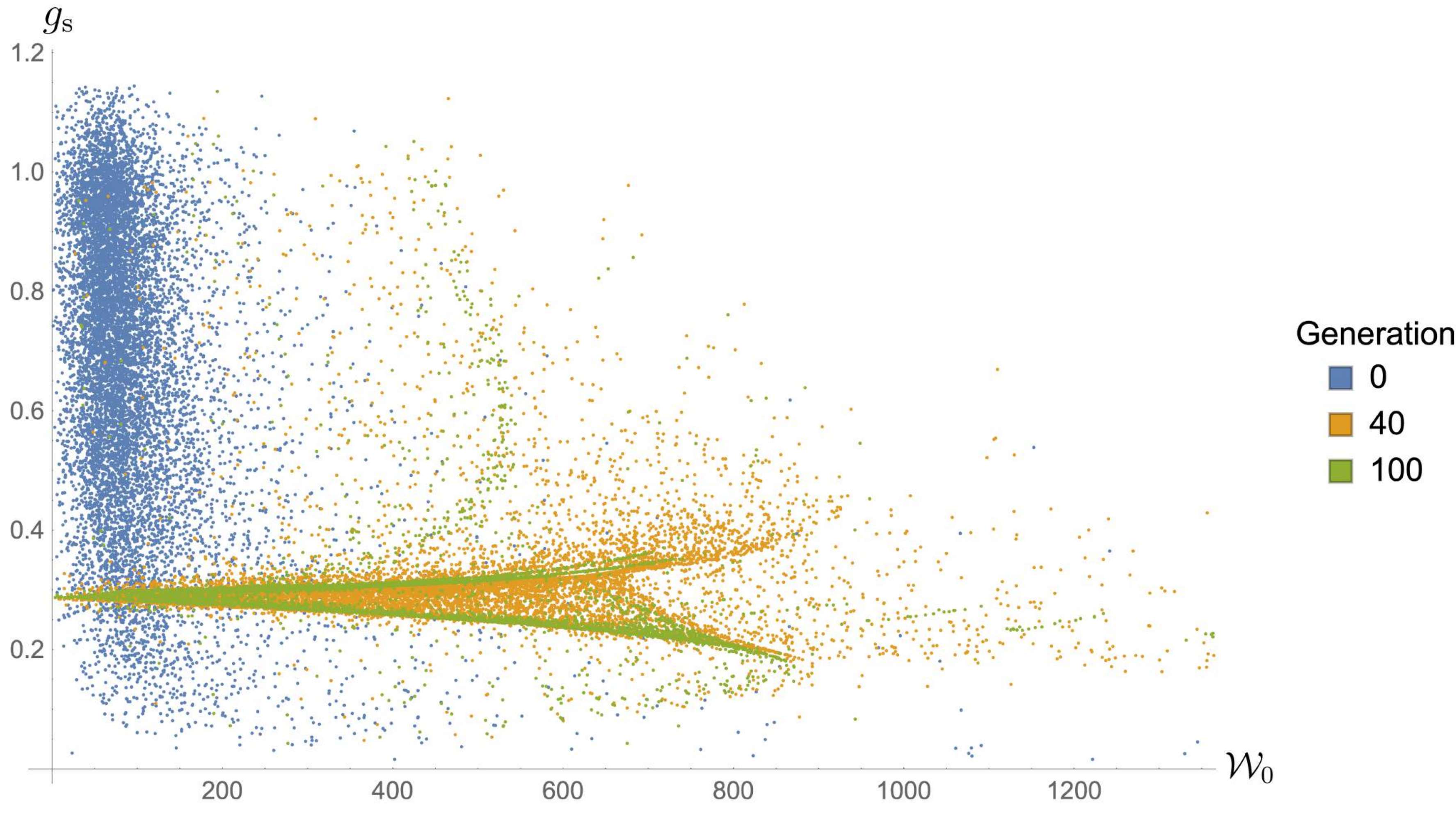}
\caption{Evolution of the distribution of $\gs$ and $\cW_{0}$ for three generations.}\label{fig:3}
\end{figure}

Using the algorithm described in Sect.~\ref{sec:GenAlg}, we find an evolution of the initial population as depicted in Figs.~\ref{fig:1} and \ref{fig:3}. Even though we only promoted $\gs$ to be a search parameter, $y=\im(\tau)$ also accumulates around a value of approximately given by $y\approx 10.68$, cf. the bottom plot of Fig.~\ref{fig:1}. As described in the introduction, this is a generic observation for GAs during which the evolution exerts a strong pull on some components of the moduli. In contrast, the values of $\re(\phi)$ and $\re(\tau)$ remain equally scattered across the fundamental domain. Remarkably, we detect a correlation between $\gs$ and $\cW_{0}$, see Fig.~\ref{fig:3}. For small values of $\cW_{0}$, the value of $\gs$ is almost perfectly pinned to $\gs^{\ast}$, whereas the solutions tend to diverge away from $\gs^{\ast}$ for large $\cW_{0}$. This clearly raises the question whether we can improve on our findings by simultaneously minimizing $\cW_{0}$. 

\begin{figure}[t!]
\centering
 \includegraphics[width=0.9\textwidth]{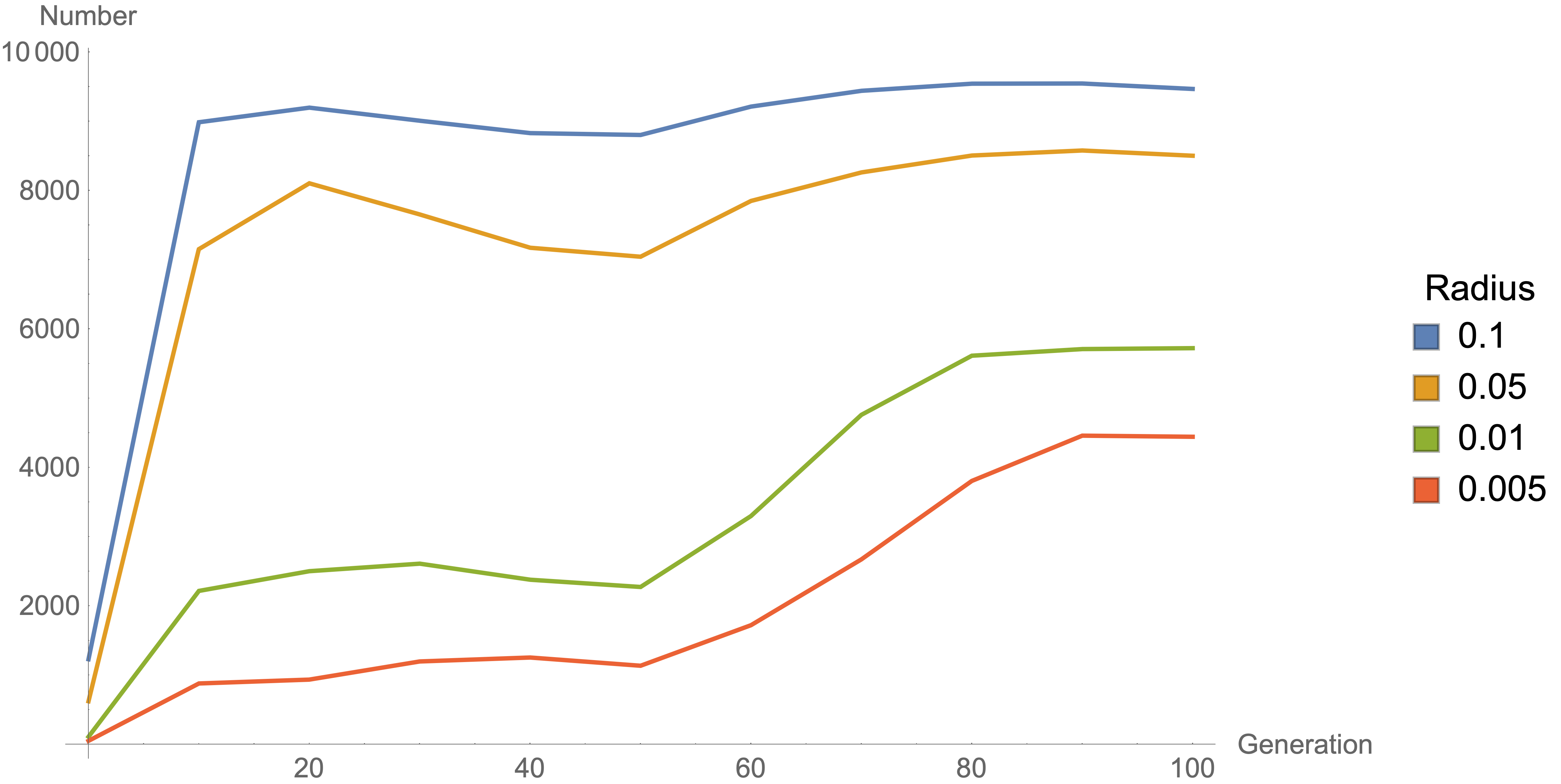}
\caption{Number of members in neighborhoods of different radius around $\gs^{\ast}$.}\label{fig:4}
\end{figure}

To quantify convergence, we count the number of members within neighborhoods of different sizes. The progress of convergence from one generation to another is shown in Fig.~\ref{fig:4}. We observe that decreasing $\delta \gs$ after generation $50$ clearly pushes the population closer to $\gs^{\ast}$. Moreover, around $90\%$ of the population can be found within the interval $[\gs^{\ast}-0.1,\gs^{\ast}+0.1]$ after only $10$ generations.

\begin{figure}[t!]
\centering
 \includegraphics[width=0.5\textwidth]{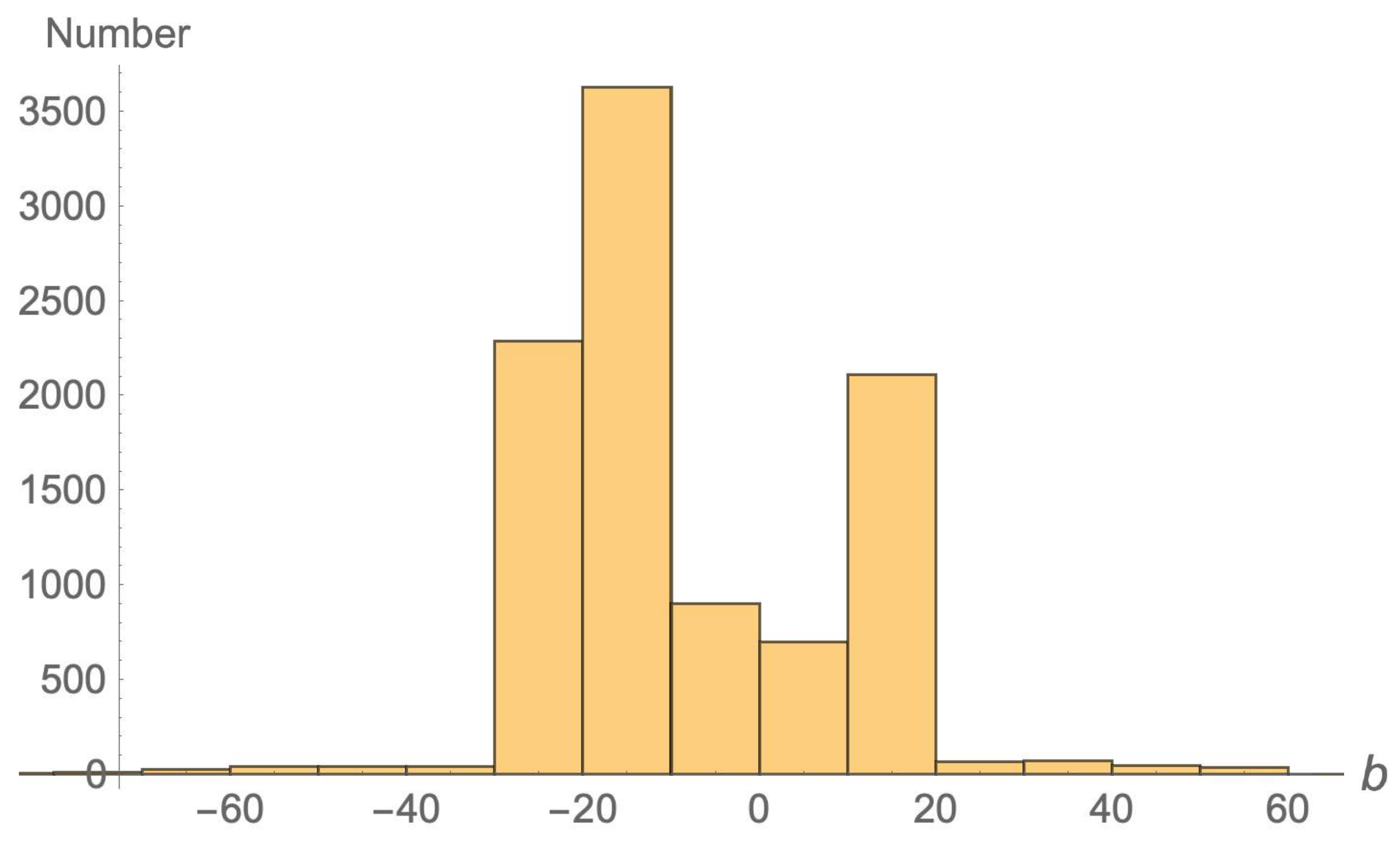}\includegraphics[width=0.5\textwidth]{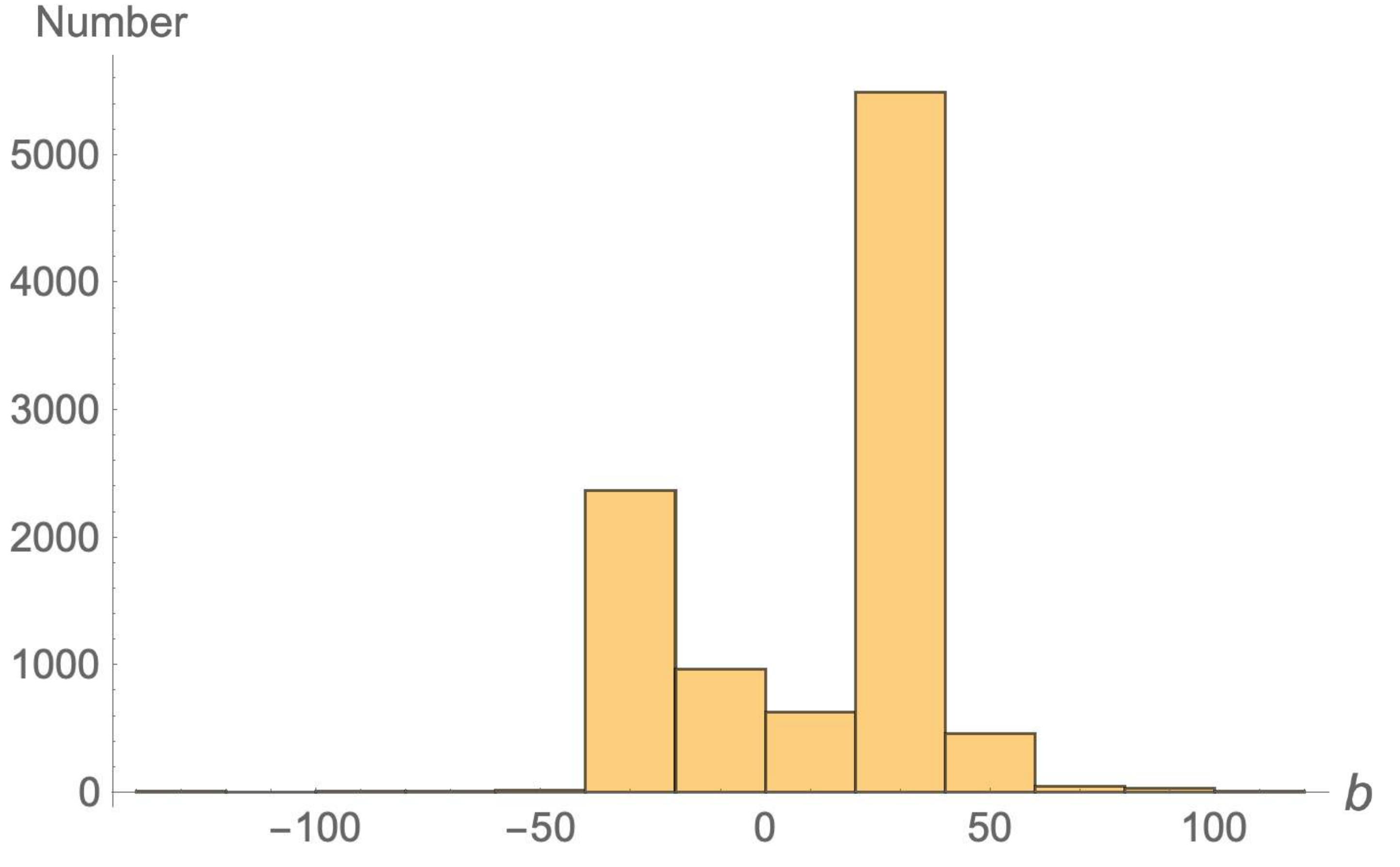}
\caption{The distribution of $b$ for generation $10$ (left) and $30$ (right).}\label{fig:T63.1}
\end{figure}

\begin{figure}[t!]
\centering
 \includegraphics[width=0.5\textwidth]{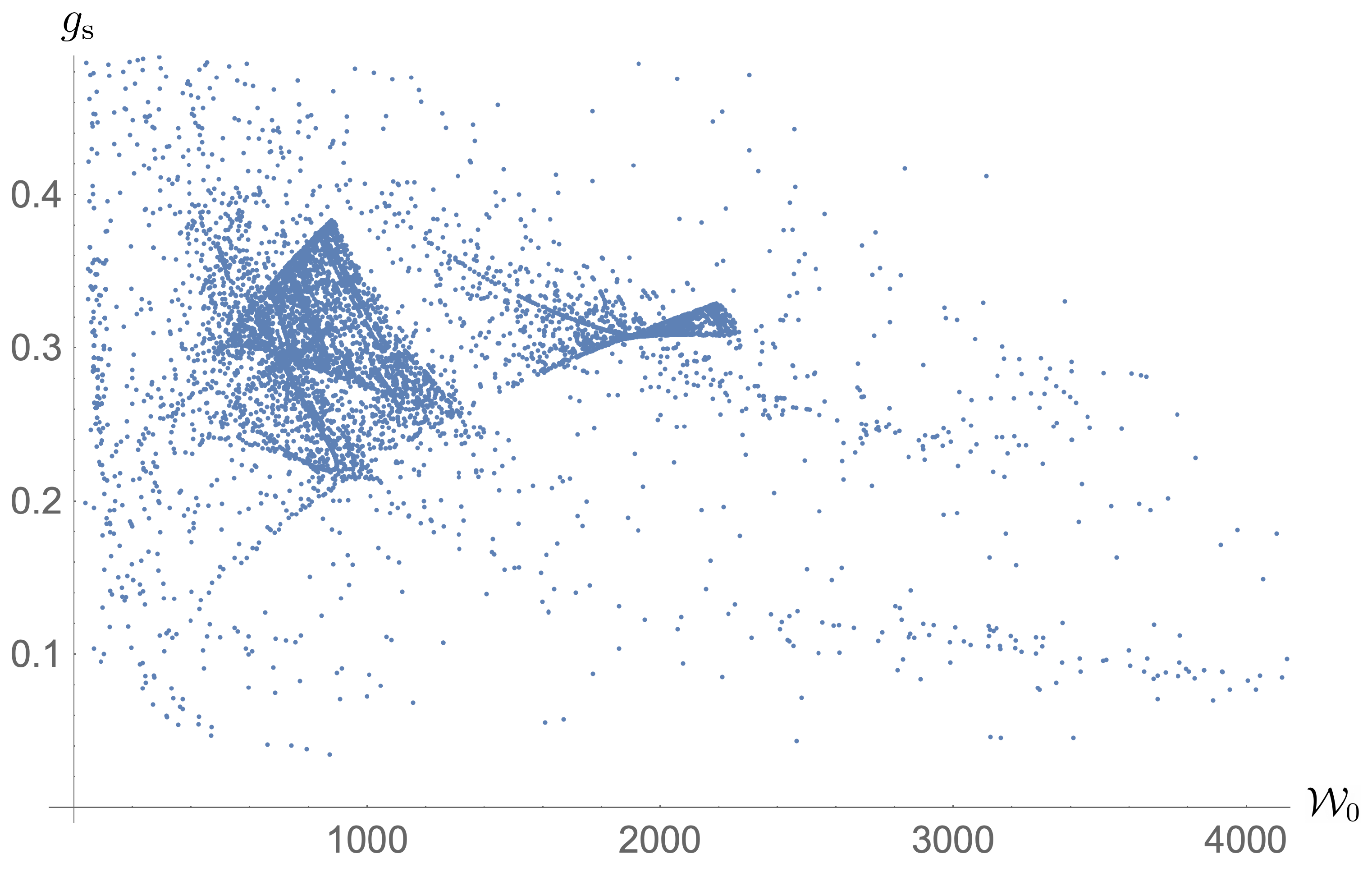}\includegraphics[width=0.5\textwidth]{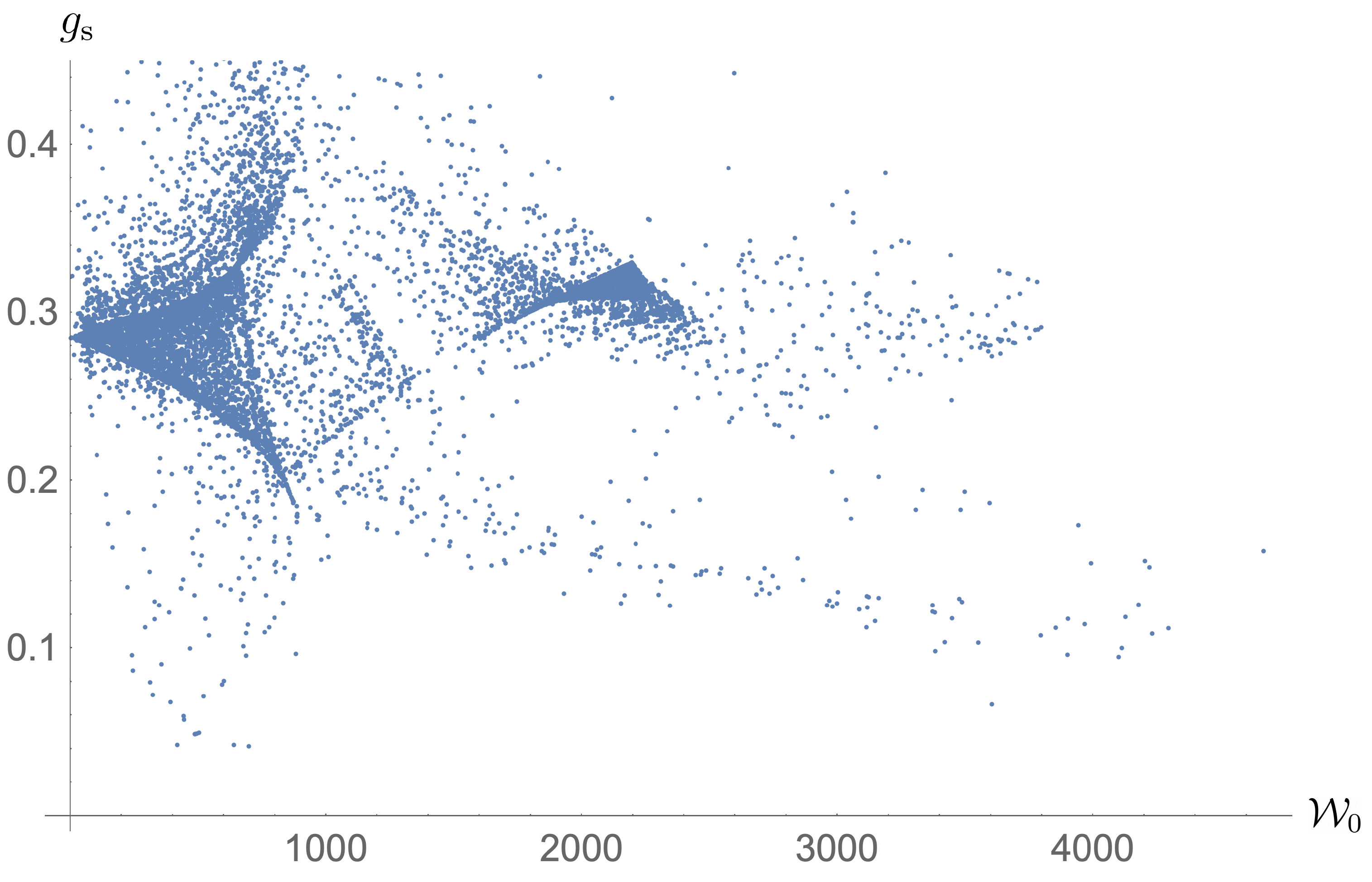}
\caption{The distribution of vacua in the $\cW_{0}$-$\gs$-plane for generation $10$ (left) and $30$ (right).}\label{fig:T63.2}
\end{figure}

As discussed in Sect.~\ref{sec:ConvEv}, the initial population evolves rapidly towards a region close to the optimal solution. The drop in blue and orange line observed in Fig.~\ref{fig:4} is associated to the fact that the population gathers around two local fitness maxima. However, the algorithm picks out a global maximum over the course of a few generations. This can also be understood by looking at the distribution of certain fluxes forming the schemata. For simplicity, we only look at one dominating flux. Fig.~\ref{fig:T63.1} shows that there are two dominating peaks in the distribution of the flux number $b$ at generation $10$ (left), recall Eq.~\eqref{eq:FluxDefT6}. Once the dominating schema wins, the population is pulled towards the peak on the right as depicted on the right hand side of Fig.~\ref{fig:T63.1}. This can also be supported by considering the distribution of vacua in the $\cW_{0}$-$\gs$-plane as shown in Fig.~\ref{fig:T63.2}. Notice that briefly the total number of members in the $1\sigma$-region decreases, cf. Fig.~\ref{fig:4}. It is therefore important to keep in mind that the purpose of the algorithm is not to find as many suitable members as possible, but to maximize the fitness function.

\begin{figure}[t!]
\centering
 \includegraphics[width=\textwidth]{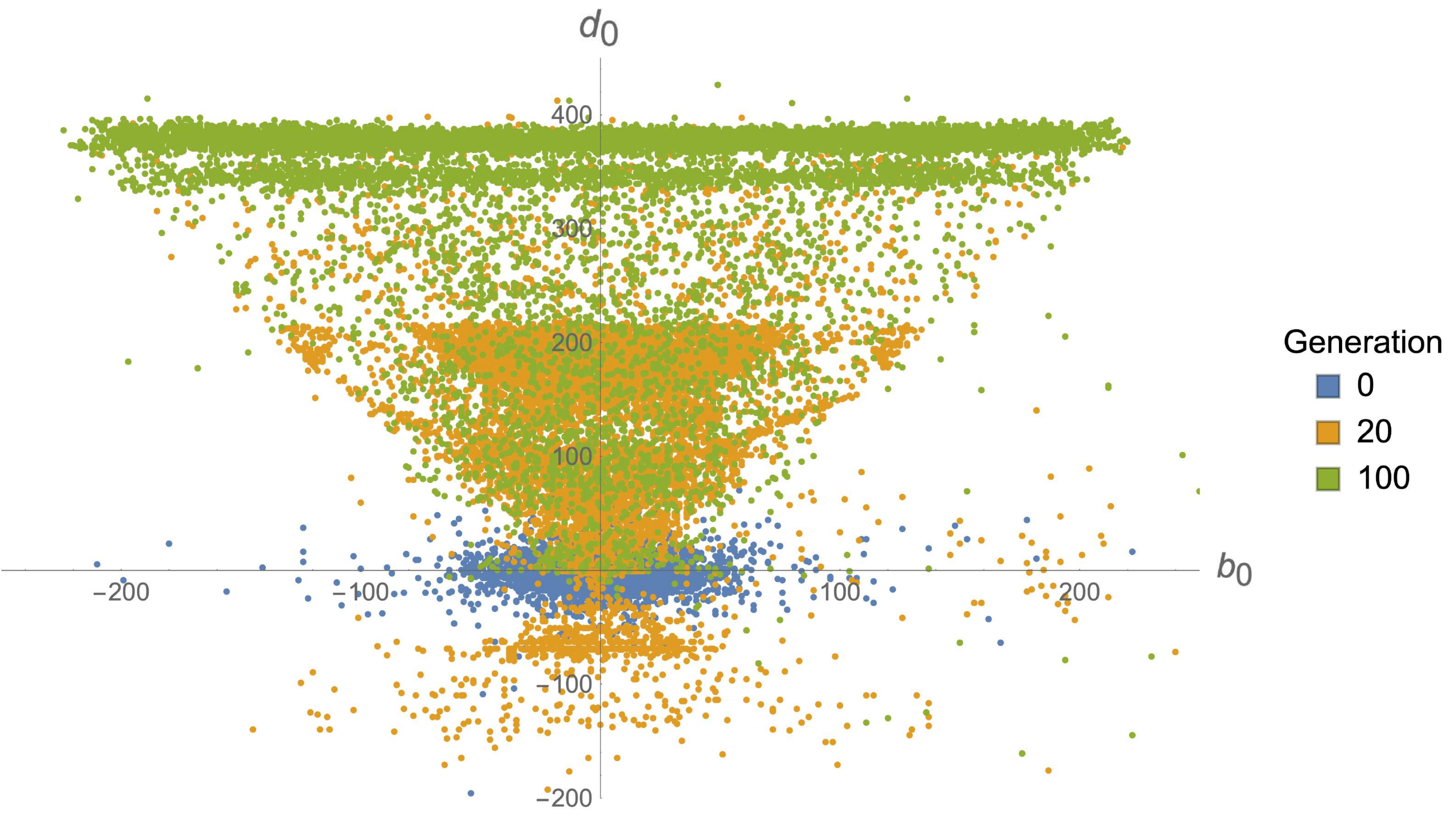}
\caption{Flux distribution for the two (almost) unconstrained fluxes $b_{0}$, $d_{0}$ in the final population.}\label{fig:T63.3}
\end{figure}

Finally, let us comment on the distribution of fluxes. By performing PCA on the fluxes we observe that there are essentially only two free unconstrained directions in flux space, namely $b_{0}$ and $d_{0}$. The evolution of the corresponding distribution is shown in Fig.~\ref{fig:T63.3}. We observe a triangular shape with two dominating horizontal lines for fixed $d_{0}$. All in all, we find that for about $80\%$ of the population
\begin{align}
&a_{0}=1\kom b_{0}\in[-250,250]\kom c_{0}=0\kom d_{0}\in [370,390]\, ,\nn\\
&a\in [-1,1]\kom b\in[-36,-33]\kom c=-1\kom d\in [-3,3]\, .
\end{align}

\subsection{Minimizing $\cW_{0}$}\label{sec:T6W0}

Here, we try to find solutions with $\cW_{0}=0$. As discussed previously, these solutions have special properties in terms of the fluxes and VEVs. Notice that we do not implement these analytic properties by hand, but simply look for solutions with $P_{1}(\tau)=P_{2}(\tau)=0$ for $\tau$ and physical VEVs.
So the algorithm does not know in advance about these special characteristic features associated to $\cW_{0}=0$ solutions. We would like to understand whether this has any impact on our genetic algorithm. We consider the parameters
\begin{equation}
\cW_{0}^{\ast}=0\kom \delta\cW=20\kom p=10000\kom N_{\text{gen}}=180\kom \lm=500\kom q_{\text{mut}}=0.8\, .
\end{equation}
We decrease the support of the fitness function by $50\%$ at generation $50$, $70$ and $90$.

\begin{figure}[t!]
\centering
 \includegraphics[width=0.9\textwidth]{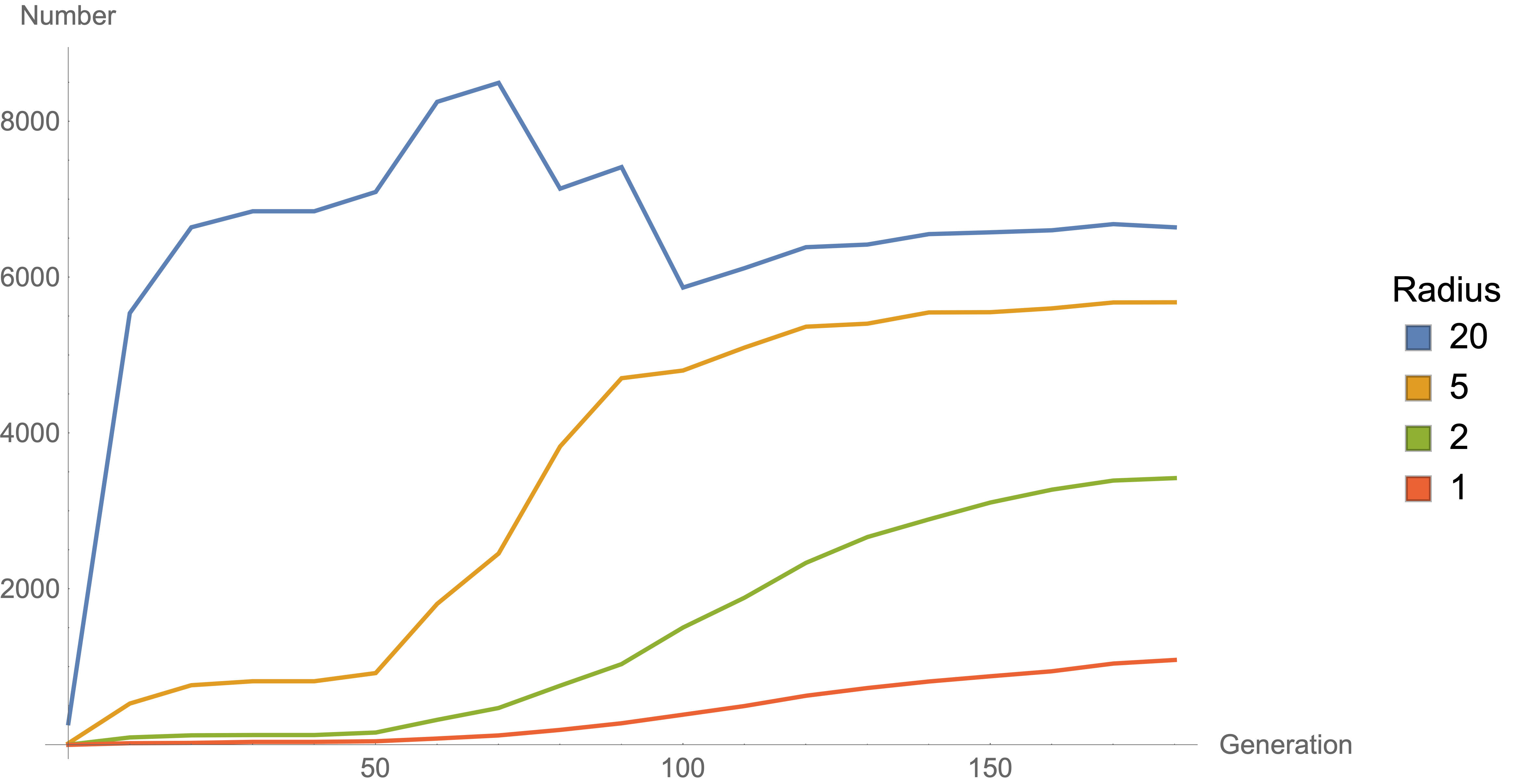}
 
 \vspace*{0.4cm}\includegraphics[width=0.9\textwidth]{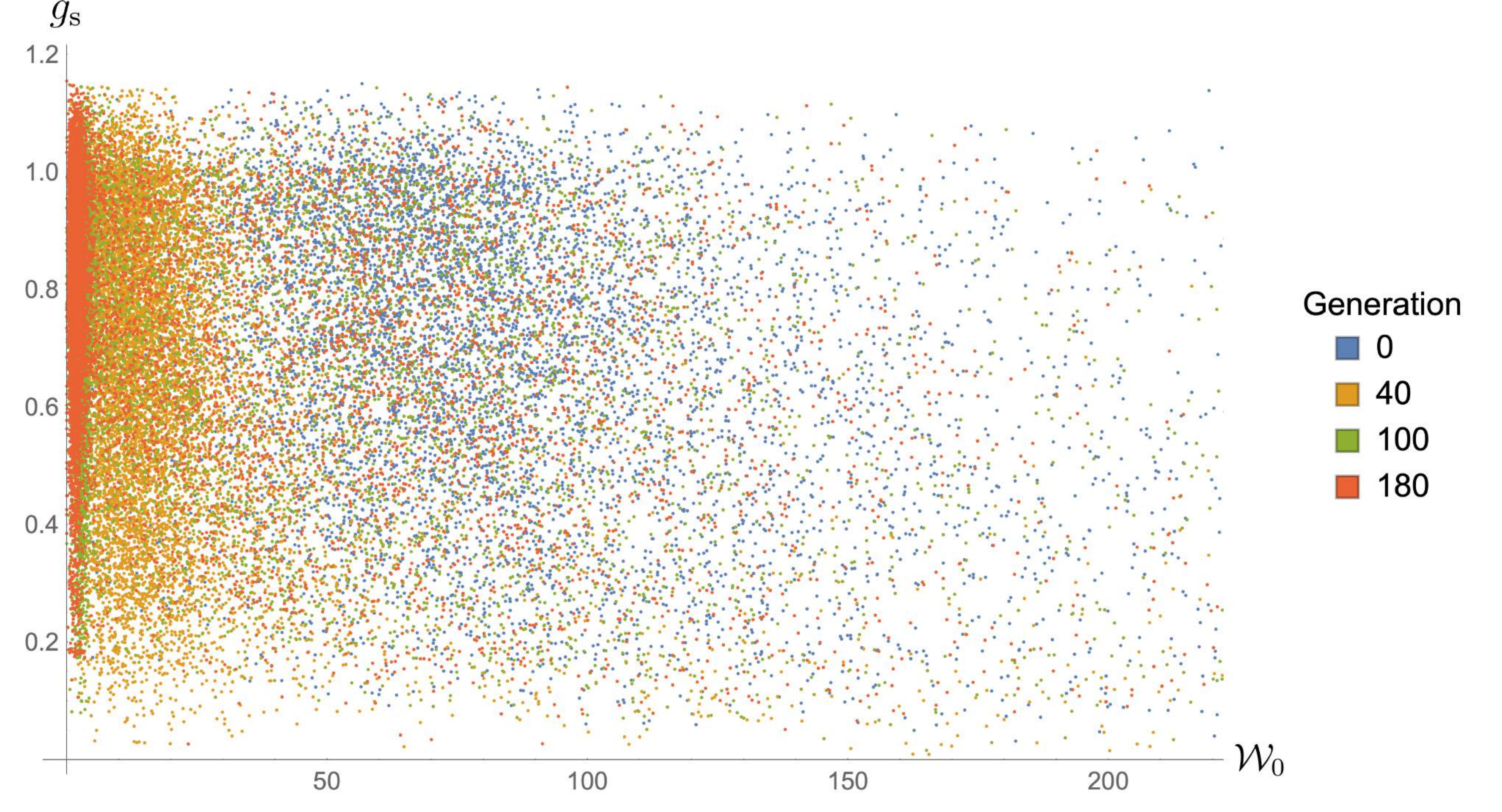}
\caption{Top: the number of members in neighborhoods of different radius around $\cW_{0}^{\ast}$. Bottom: the evolution of the distribution of $\gs$ and $\cW_{0}$ for four generations.}\label{fig:N10}
\end{figure}

\begin{figure}[t!]
\centering
 \includegraphics[width=0.6\textwidth]{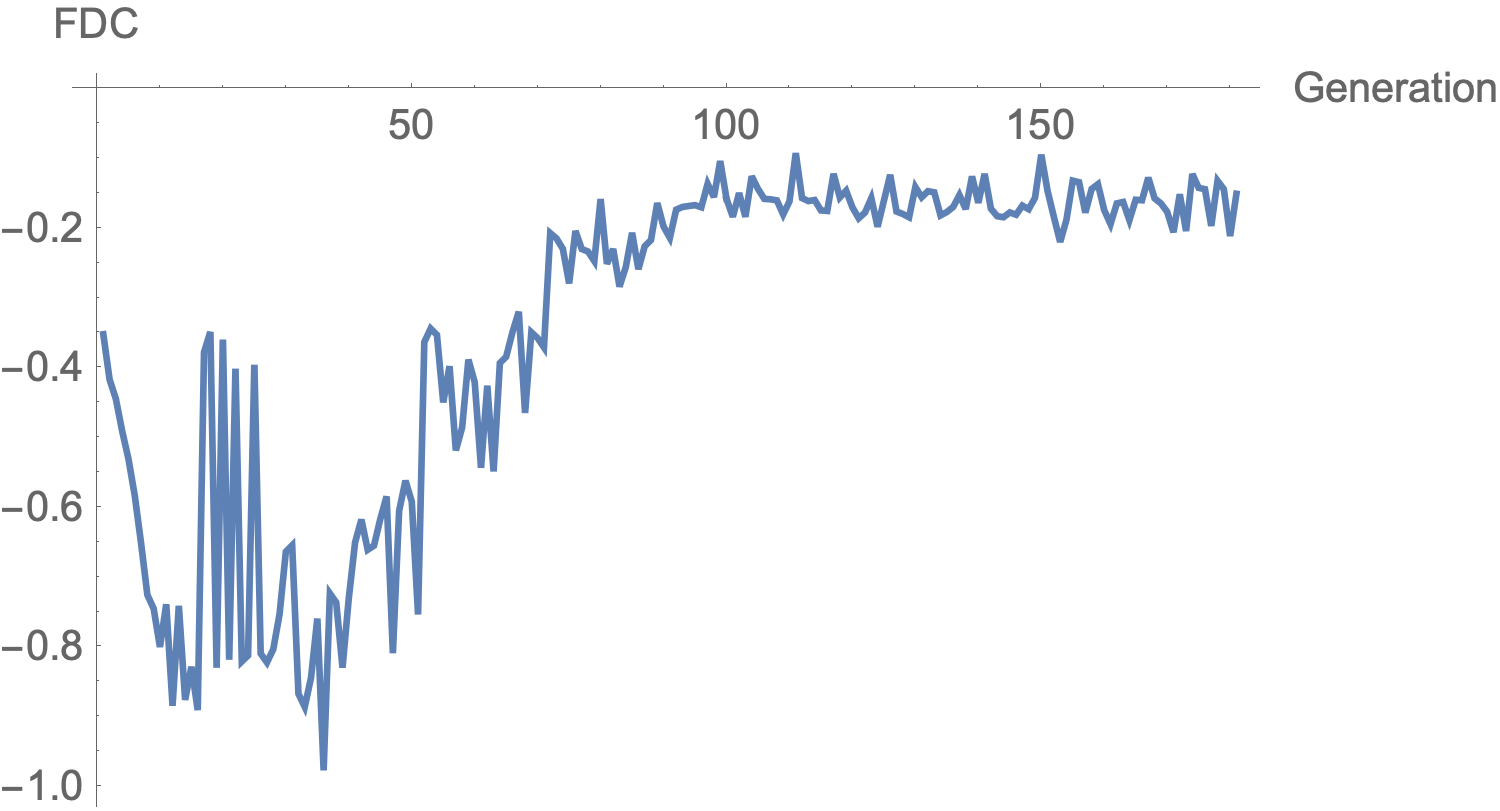}
\caption{Fitness distance correlation as a function of generation.}\label{fig:N11}
\end{figure}

As shown in Fig.~\ref{fig:N10}, the populations quickly converge towards small values of $\cW_{0}$. However, the rate of convergence is significantly less than in e.g. Fig.~\ref{fig:4}. This becomes even more apparent when investigating the FDC of this task. Indeed, the evolution of the FDC in Fig.~\ref{fig:N11} shows that, while minimizing itself is not problematic at all, finding $\cW_{0}<2.5$ is GA-difficult. For one thing, the lack in convergence is related to the scarcity of $\cW_{0}=0$-solutions in comparison to those with $\cW_{0}\neq 0$. As derived in \cite{DeWolfe:2004ns}, the number $N_{\text{vac}}(L,\cW_{0}=0)$ of $\cW_{0}=0$-solutions for given flux scale $L\leq \lm$
\begin{equation}
\dfrac{N_{\text{vac}}(L,\cW_{0}=0)}{N_{\text{vac}}(L)}\sim \dfrac{9}{8\pi^{2}}\dfrac{\log(L)}{L}
\end{equation}
in terms of the total number of vacua $N_{\text{vac}}(L)$. On top of that, at small $L$ discretization effects overshadow $\cW_{0}=0$ solutions (see for example Fig.~3 in \cite{DeWolfe:2004ns}) which is why we have to go to comparatively large $L$. In any case, we observe that special solutions with $\cW_{0}=0$ are difficult for our algorithm to identify. Notwithstanding, we note that a small fraction of individuals do satisfy $\cW_{0}=0$.

In the light of the ongoing discussion about KKLT and related dS-vacua constructions \cite{Obied:2018sgi,Ooguri:2018wrx}, it is crucial to search the landscape for suitable flux vacua, see e.g. \cite{Giryavets:2003vd,Denef:2004dm} for previous attempts. In order to safely ignore perturbative corrections, it is necessary to have $\cW_{0}\ll 1$ \cite{Kachru:2003aw,Conlon:2005ki}. One can infer the existence of such vacua by employing statistical arguments \cite{Denef:2004ze}. However, it is generally true that these vacua are less common than those with $\cW_{0}\gtrsim \cO(1)$ using similar arguments to the one above \cite{Denef:2004cf}.

In our case, the smallest value found via the above algorithm is $\cW_{0}\sim 10^{-2}$. Here it is important to keep in mind that the symmetric $T^{6}$ is only a simple toy example. General Calabi-Yau orientifolds with many moduli are expected to have a more involved structure within their flux landscape. Therefore, it would be interesting to see whether smaller values of $\cW_{0}$ are obtainable within such geometries utilizing GAs and extensions thereof.

\subsection{Fixing mass scales}\label{sec:masses}

To conclude our survey of applications, we investigate a more difficult problem in the context of GAs. Namely, we elaborate on the idea of specifying several observables within our search. As we will see further below, it is necessary to try different selection methods in order to guarantee the success of our GA. We restrict our analysis to the case of no crossover which allows us to study the impact of different selection methods more easily.

The question of finding flux configurations with certain values of the mass scales is ubiquitous in the context of model building in string theory. For simplicity, we compute the diagonal entries of the mass matrix \eqref{eq:MassMatrix} for the canonically normalized fields evaluated at the SUSY minimum $D_{\phi}\cW=D_{\tau}\cW=0$ which we denote $M_{\re(\phi)}$ etc. We fix all four masses at the same value, i.e., we consider
\begin{equation}
M^{\ast}=5000\kom \delta M=1000\kom p=10000\kom N_{\text{gen}}=120\kom q_{\text{mut}}=1\, .
\end{equation}
We determined the associated weights and the offset in \eqref{eq:GenFitness} by trial and error, giving
\begin{equation}
w_{\re(\phi)}=6.8\kom w_{\im(\phi)}=1.0\kom w_{\re(\tau)}=1.3\kom w_{\im(\tau)}=0.4\kom b=0\, .
\end{equation}
We decrease $\delta M$ only once at generation $60$.

\begin{figure}[t!]
\centering
 \includegraphics[width=0.7\textwidth]{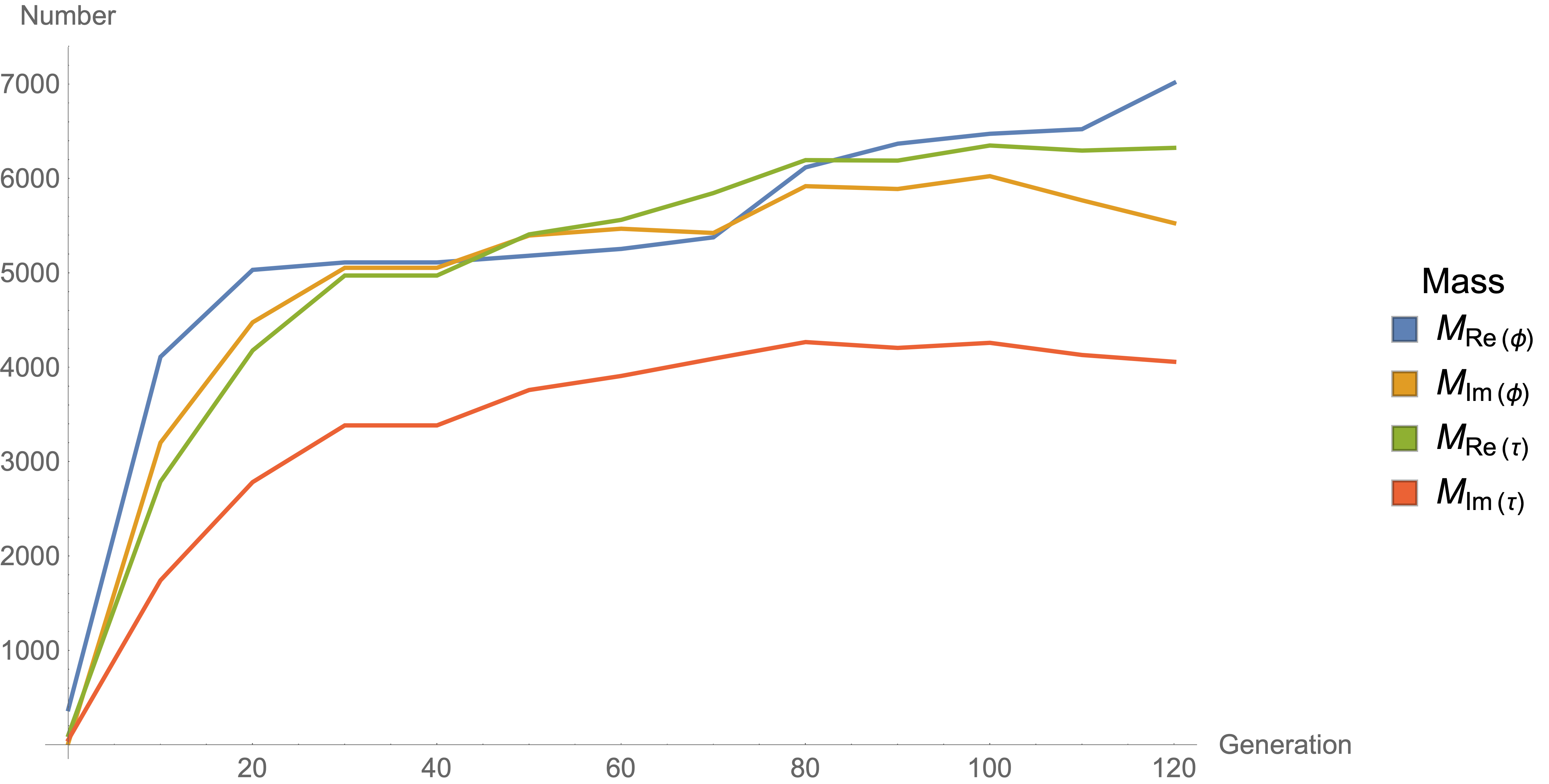}

\vspace*{0.4cm} 
 \includegraphics[width=0.7\textwidth]{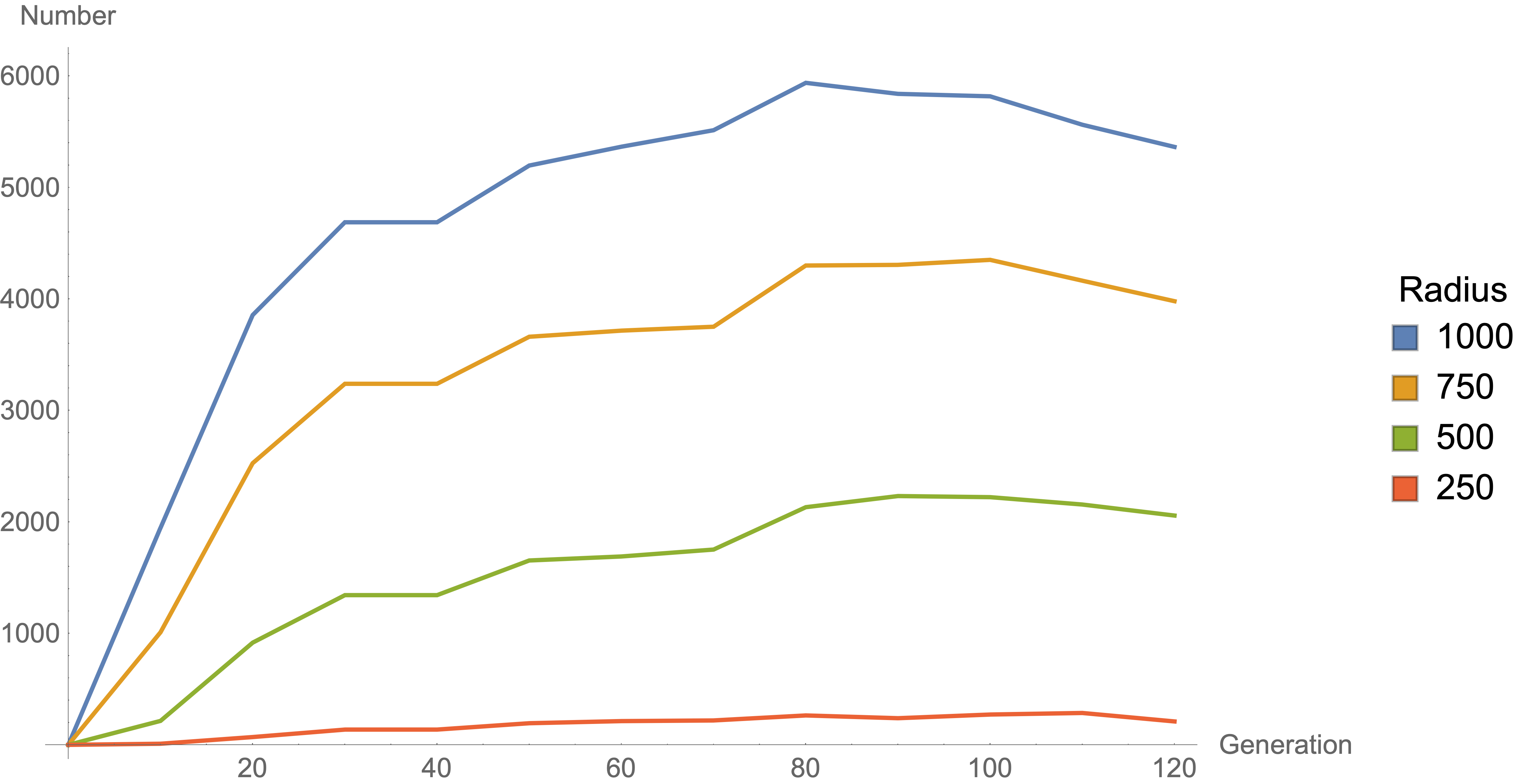}
\caption{Top: the number of individuals having a single mass matrix entry within a neighborhood of $r=500$. Bottom: the number of individuals having all four masses within neighborhoods of different radii.}\label{fig:Mass}
\end{figure}

Figure \ref{fig:Mass} summarizes the results obtained by running the GA. The left plot shows that certain quantities are treated differently from the others during the algorithm's evolution. The right hand side of Fig.~\ref{fig:Mass} depicts the number of individuals that satisfy the search criteria for all four mass scales at the same time. In comparison to the results obtained in the previous section, we conclude that it is significantly harder to get closer to the optimal solution whenever several parameters are taken into account. However, this might simply be related to a bad choice of selection methods. As mentioned previously, many of such methods have been discussed in the literature, but we restricted mainly to the so-called \emph{roulette-wheel} selection so far where the fitness itself is used as a measure of probability to procreate. This typically leads to premature convergence towards a local fitness maximum. This is not really problematic in the case of one GA-parameter. In the context of several, however, this is a serious issue.

\begin{figure}[t!]
\centering
 \includegraphics[width=0.7\textwidth]{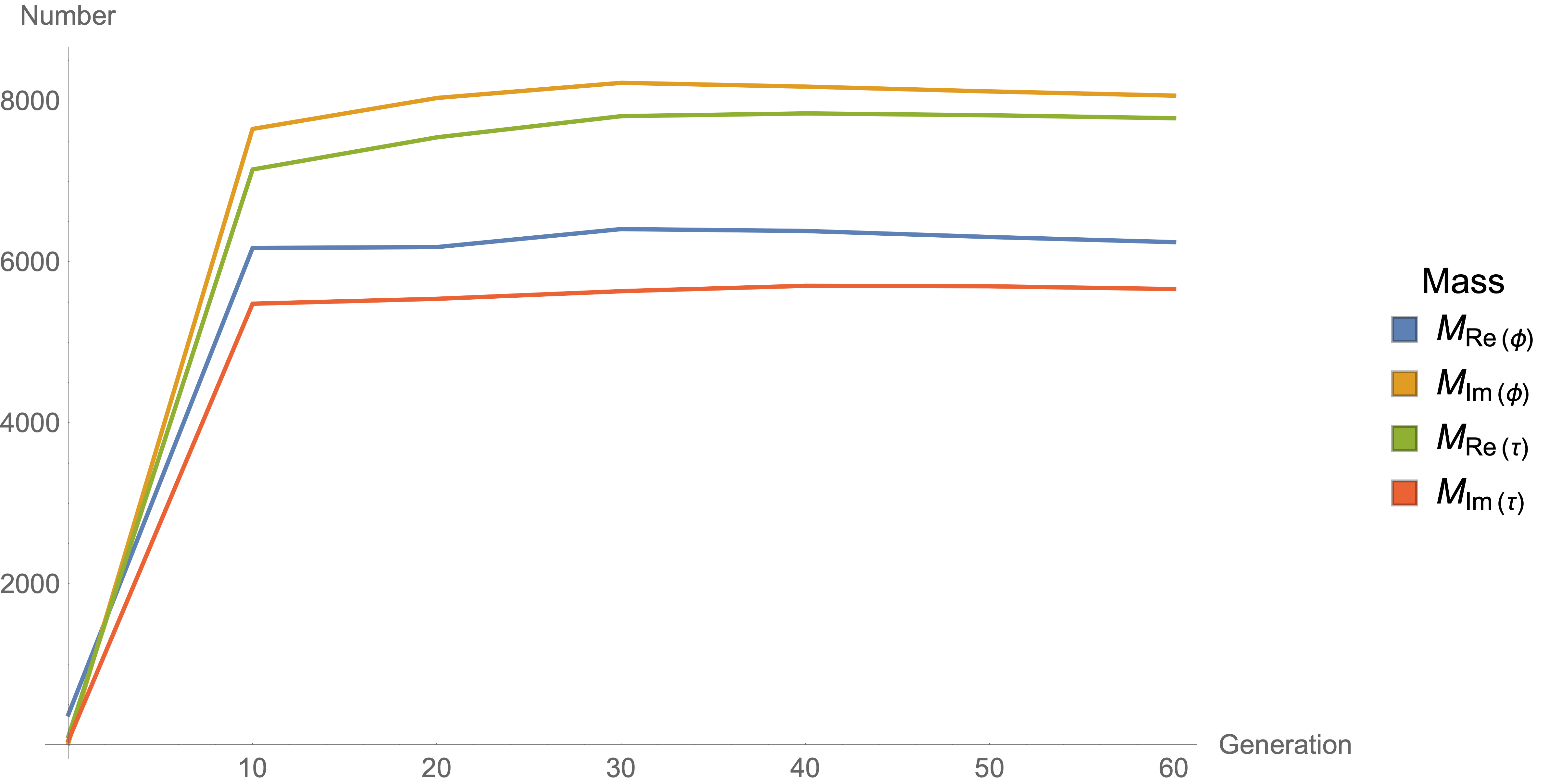}
 
\vspace*{0.4cm} \includegraphics[width=0.7\textwidth]{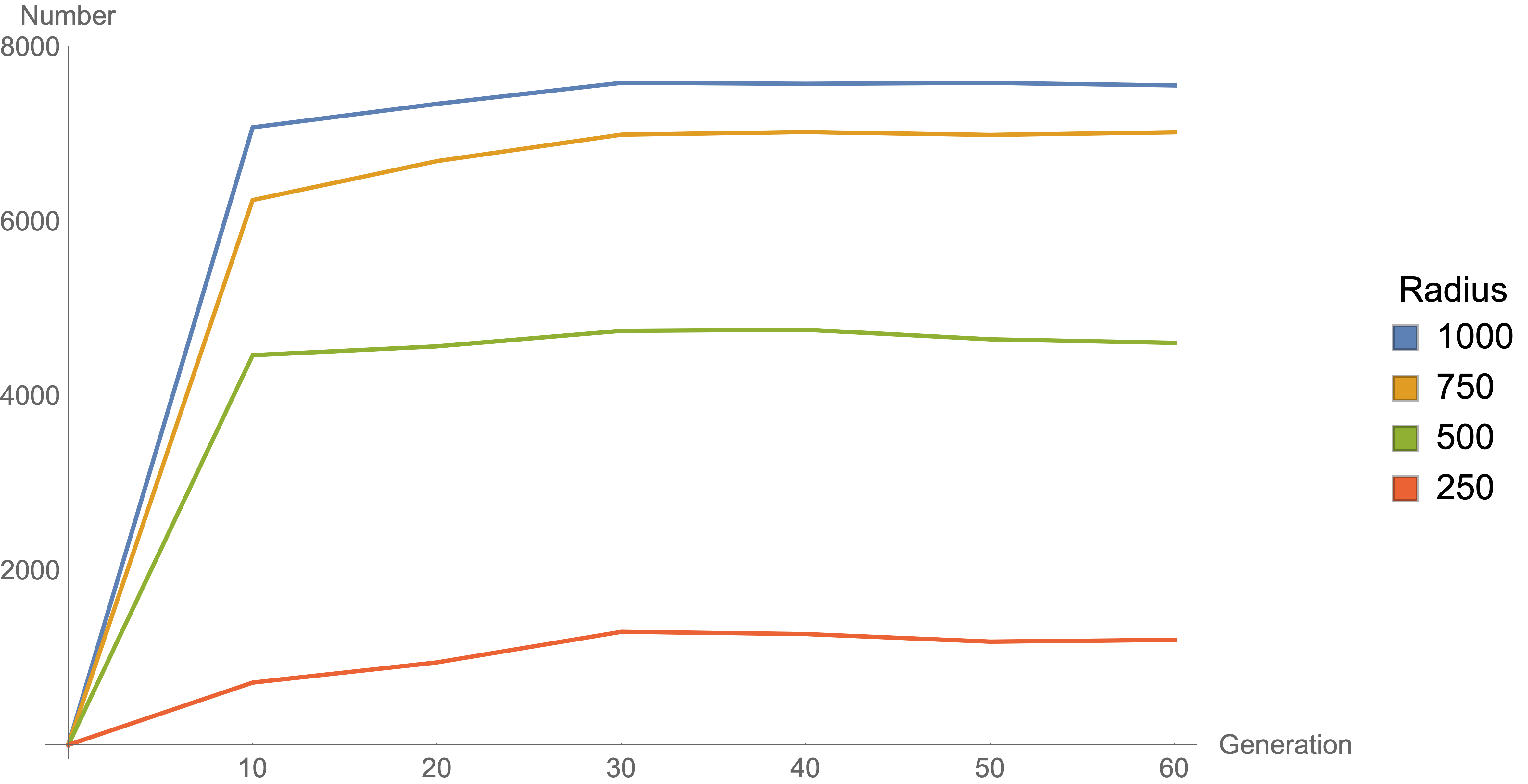}
\caption{Top: the number of individuals having a single mass matrix entry within a neighborhood of $r=500$ using tournament selection. Bottom: the number of individuals having all four masses within different neighborhoods.}\label{fig:MassTour}
\end{figure}

Therefore, we tested different selection techniques such as 
\begin{itemize}
\item \emph{rank-weighted selection} where each member is ranked by a number $1,\ldots,p$ according to its fitness. This ranking is used to determine the probability for procreation which is chosen to be a linear mapping here. In comparison to roulette-wheel selection this means that individuals with high fitness do not dominate the crossover procedure too early during evolution.
\item \emph{tournament selection} where so-called tournaments between a subset of members are performed. That is, we pick $k<p$ individuals from our population and take only the fittest for breeding. Since we need to make several of such tournaments, the diversity of fit individuals in the descendant population is enhanced. In other words, the population does not localize too early around a local fitness maximum.
\end{itemize}
The first does not lead to any better results, but is significantly slower in terms of convergence around the optimal solution. In contrast, tournament selection is really efficient in determining a global fitness maximum that satisfies the constraints on all search parameters.

Starting with the same initial population as above, we use the parameters
\begin{equation}
w_{\re(\phi)}=w_{\im(\phi)}=w_{\re(\tau)}=w_{\im(\tau)}=1.0\kom b=0
\end{equation}
and consider only $N_{\text{gen}}=60$ generations. The results for tournament selection are summarized in Fig.~\ref{fig:MassTour}. A crucial advantage of using tournament selection is that one can achieve significantly better results compared to Fig.~\ref{fig:Mass} with only half the number of generations. Even though performing several tournaments is computationally more expensive, this selection method is therefore more efficient in the scenario under consideration. 

{Finally, we would like to discuss tasks that are hard to tackle with GAs. In the context of this section, we observed some difficulty trying to arrange mass hierarchies. We believe that this is related to the more general circumstance that, whenever observables are strongly correlated, it is hard to fix them simultaneously. In the worst case, certain parameter values might be incompatible with each other, in the sense that no such pair exists in the fitness landscape. Alternatively, such solutions are very scarce. Identifying these limitations is typically beyond the GA's (or any stochastic search algorithm's) capabilities. 
%However, taking these correlations and forbidden regions in moduli space into account could alleviate this issue. We leave this an open question to be discussed in upcoming projects.
}

%%%%%%%%%%%%%%%%%%%%%%%%%%%%%%%%%%%%%%%%%%%%%%
\section{Conclusions}
\label{sec:summary}
%%%%%%%%%%%%%%%%%%%%%%%%%%%%%%%%%%%%%%%%%%%%%%%%

In this paper, we demonstrated that genetic algorithms are a helpful tool for searching for specific phenomenological features in the landscape of flux vacua. This is possible due to the underlying structure of the landscape, which is crucial for the success of GAs. We showed that GAs may be used not only to speed up the search for phenomenologically interesting models, but also to
uncover unknown structures and correlations in the landscape.

In the present work, we discussed various properties of GAs that allow for a systematic and efficient study of the flux landscape. Specifically, we proposed a dictionary between concepts in these two respective fields that could be useful for
 future applications of GAs to string theory. 

Applying our dictionary, we first considered the example of a Calabi-Yau hypersurface in a weighted projective space close to a conifold point. 
We observed the important effect the mutation rate has on the algorithm's performance, and the need to adjust it for various applications. 
We also studied the general dynamics of the GA, with an initial period of strong pull towards a local optimum followed by slower convergence to the global optimum.
Along these lines, we used TDA and PCA to study the ``shape'' of the distribution of fluxes as a function of generation. These methods confirmed our ideas regarding clustering in the final population.
{We showed that distinct crossover operations typically lead to different evolutionary properties of the GA, albeit the final population turns out to be almost equivalent.} 
{In addition,} we compared {various} breeding mechanisms which could be useful in scaled-up versions of the models investigated in this paper. In contrast to a Metropolis algorithm, the GA identifies links between various vacua due to the applied crossover procedure, enabling it to exploit the landscape's structure. On the other hand, a Metropolis algorithm features no such benefits.

We also applied the GA to a symmetric $T^{6}$. While searches for values of the string coupling are GA-easy, we showed that finding solutions satisfying $\cW_{0}=0$ is difficult in the context of GAs. Although these solutions emerged within our searches, their scarcity seems to counteract the formation of associated schemata.
This behavior can be quantified in terms of the fitness distance correlation. Finally and most importantly, we demonstrated that GAs are capable of finding vacua with several properties, namely specifying several masses. This success is encouraging as we look towards phenomenologically-viable constructions.

We emphasize that the applications of GAs are diverse and can easily be translated into other promising setups of, e.g., model building in string theory. In the future, we would like to turn to phenomenologically interesting questions such as testing the WGC or other statements about the landscape. We would also like to investigate different selection methods and other GA-relevant techniques allowing for an improved version of the algorithm applied in this paper.

{\bf Acknowledgments}

We thank Arthur Hebecker and Pablo Soler for discussion.
This work is supported in part by the DOE grant DE-SC0017647 and the Kellett Award of the University of Wisconsin.
AC would like to thank the Straka Fund at UW-Madison for financial support.
GS would like to thank 
the Aspen Center for Physics for hospitality during the final stages of this work.
AS acknowledges support by the German Academic Scholarship Foundation and the hospitality of the University of Wisconsin during part of this work.

\bibliographystyle{JHEP}
\bibliography{Literatur}

\end{document}